\documentclass[pra,reprint,superscriptaddress]{revtex4-2}
\usepackage{epsfig}
\usepackage{graphicx}
\usepackage{amsmath}
\usepackage{amssymb}
\usepackage{mathrsfs}
\usepackage{dcolumn}
\usepackage{amsfonts}
\usepackage{ulem}
\newcommand{\Tr}{\text{Tr}}
\newcommand{\ket}[1]{|#1\rangle}
\newcommand{\bra}[1]{\langle#1|}

\begin{document}
\title{Influence of initial states on memory effects: A study of early-time superradiance}
\author{S. C. Hou}
\email{ housc@dlmu.edu.cn}
\affiliation{School of Science, Dalian Maritime University, Dalian 116026, China}
\author{G. Q. Shuai}
\affiliation{School of Science, Dalian Maritime University, Dalian 116026, China}
\author{X. Y. Zhang}
\affiliation{School of Science, Dalian Maritime University, Dalian 116026, China}
\author{J. Shen}
\affiliation{Department of Criminal Technology, Liaoning Police College, Dalian 116036, China}
\author{X. X. Yi} 
\email{ yixx@nenu.edu.cn}
\affiliation{Center for Quantum Sciences and School of Physics, Northeast Normal University, Changchun 130024, China}
    
\begin{abstract}
The initial states of a quantum system can significantly influence its future dynamics,
especially in non-Markovian quantum processes due to the environmental memory effects.
Based on a previous work of ours, we propose a method to quantify the memory effects of
 a non-Markovian quantum process conditioned on a particular system initial state. 
With this method, we analytically study the early-time memory effects of 
a superradiance model consisting of $N$ atoms (the system) interacting with a 
single-mode vacuum cavity (the environment) with two types of initial states:  
the Dicke states and the factorized identical states.  We find that the radiation
intensity are closely related to the memory effects, and correspondingly, the enhancement
of memory effects (from independent radiation to collective radiation) is important
for the degree of superradiance, especially for the Dicke states. Furthermore, numerical 
results in other regimes show that the characteristics of memory effects and superradiance
in longer-time dynamics can be reflected through its early-time dynamics.  
\end{abstract}
\date{\today}
\maketitle

\section{Introduction}
The initial state of a quantum system can significantly influence its future dynamics \cite{Cordero, Wenderoth},
especially in non-Markovian quantum processes due to the environmental memory effects.
One trivial example is that if the system is initially in a steady state in a non-Markovian
quantum process, it can hardly exhibit any non-Markovian features afterward,
such as the nonmonotonic behaviors of energy and information flows \cite{Rivas2014,Breuer2016,Vega2017,Li2018}.
More intriguing phenomena may emerge when a system consists of a collection of subsystems, 
such that the properties of its initial state, such as entanglement and coherence,
 may significantly influence its future dynamics. A well-known example is the concept of superradiance 
\cite{Dicke,Rehler,Andreev,Gross,Agarwal1970,Bonifacio.I,Bonifacio.II,Carmichael} introduced
 by Dicke in 1954, where the emission intensity from an ensemble of atoms interacting with a 
 common electromagnetic field can be enhanced compared with that from independent atoms.  
 It is also well-known that the superradiance behaviors are highly relevant to the initial 
 states of the system. In recent years, superradiance has received a
 large amount of attention due to its theoretical significance and potential applications 
 \cite{Scully2009,Kim, Mlynek,Rastogi,Robicheaux,Mogilevtsev,Bojer,Paulisch,Cheng,
 Tan,Wolfe,Tasgin,Lohof,Dinc,Sinha,Qiu2023,Zhang2023}. Under certain approximations, 
 such as a coarse-grained timescale, the superradiance process could be 
regarded as Markovian \cite{Gross,Agarwal1970,Bonifacio.I,Bonifacio.II,Carmichael}, whereas,
it is intrinsically non-Markovian. With the advances in theories and technologies, 
understanding the non-Markovian dynamics of superradiance becomes more demanding 
\cite{Dinc,Sinha,Qiu2023,Zhang2023}.  In a recent work \cite{Zhang2023}, the author shows
that non-Markovian memory effects play an important role in superradiance beyond retardation, 
featuring the quadratic dynamics in the early-time (Zeno) regime, whereas, 
the memory effects are not quantitatively evaluated. In view of the significant influences
of the system initial states on non-Markovian quantum processes, especially, a superradiance process, 
some interesting questions arise.  For example, how to quantitatively evaluate the memory 
effects of a quantum process conditioned on a particular system initial state? 
Are there quantitative relations between the memory effects and the superradiance 
characteristics in the early-time  regime? What is the role of the (very weak) 
memory effects in a superradiance process that could be well approximated as Markovian?

In recent years, a number of measures or manifestations of non-Markovianity were 
proposed to characterize the memory effects which are often connected to nonmonotonic behaviors \cite{Rivas2014,Breuer2016,Vega2017,Li2018}. For example, the well-known BLP 
measure \cite{Breuer2009} and the RHP measure \cite{Rivas2010} use the increases of distinguishability 
and entanglement, respectively, to measure the non-Markovianity. The nonmonotonic behaviors 
originate from the noncomplete positivity of the intermediate dynamical maps in non-Markovian processes. 
These measures generally apply to quantum processes where the evolutions start at a fixed time $t_0$ 
(usually $t_0=0$ for simplicity). However, a key distinction of the non-Markovian 
process is the existence of $t_0$ in the generator of the system's time-local equation, i.e., 
$\dot{\rho}_S=\mathcal{L}(t-t_0)\rho_S$ \cite{Chruscinski}. Thus some features of 
non-Markovianity might not be characterized with only evolutions starting from
a fixed time $t_0$. For example, in the early-time regime or a near-Markovian regime, 
the spontaneous radiation of an atom could be described by a time-local equation
with a positive (time-dependent) damping rate, which looks like a time-dependent Markovian 
master equation (except $t_0$ in its generator) \cite{Hou2015}. Thus the nonmonotonic behaviors
do not occur in these regimes, implying weaker memory effects. Meanwhile, the 
non-Markovianity measures based on the noncomplete positivity may give zero 
memory effects. On the other hand, the previously proposed non-Markovianity measures 
generally do not depends on a particular system initial state, but deal with a 
quantum process where all the potential initial states are considered. 
Therefore, an initial-state-dependent measure which is able to characterize 
weak memory effects is demanding. 

 In this paper, we propose such a method developed from a previous work \cite{Hou2015} of ours. 
The non-Markovianity measure in Ref. \cite{Hou2015} is based on the inequality of completely positive 
dynamical maps $T(t_2,t_0)\neq T(t_2,t_1)T(t_1,t_0)$ and connected with the validity of the
Markov approximation. By acting the inequality on a particular initial state $\rho_S(t_0)$, 
the evolutions starting at $t_0$ and $t_1$ have the same system states at $t_1$ but different histories before $t_1$. 
Thus the memory effects (past-future dependence) could be understood by the difference between
 the final states $\rho_S(t_2)$ and $\rho_S'(t_2)$.  Based on the above interpretation, 
 we suggest measuring the memory effects conditioned on the initial state $\rho_S(t_0)$ by 
 the maximal difference of $\rho_S(t_2)$ and $\rho_S'(t_2)$  in a time interval of interest.  
 Using this method, we calculate the memory effects (as well as the radiation characteristics) 
 of a superradiance model with two types of initial states: the Dicke states and 
 factorized identical states in different regimes. The model describes $N$ two-level atoms 
 interacting with a single-mode cavity initially in vacuum. We give analytic expressions of 
 the characteristics of memory effects (and superradiance) in the early-time regime. 
 Moreover, we extend our work to longer time intervals in the near-Markovian regime,
and to the whole radiation lifetime in a strongly non-Markovian regime with numerical results.

The main findings are as follows. 1. The radiation intensity is closely related to the memory effects,
 and correspondingly, the degree of superradiance is closely related to the degree of memory-effect-enhancement 
(from independent radiation to collective radiation), especially for the Dicke states. 
2. The entanglement in the Dicke states and the single-atom coherence in the factorized 
identical states are necessary for the superradiance and important for the memory-effect-enhancement.  
3. The (change of) environmental photon number is a main physical source of memory effects 
for our model. 4. The characteristics of memory effects and superradiance in long-time 
dynamics (especially for the near-Markovian regime) can be reflected through its early-time dynamics.

The paper is organized  as follows: In Sec. \textrm{II},  we propose our method to evaluate 
the initial-state-dependent memory effects in a quantum process. 
In Sec. \textrm{III}, we obtain the early-time solution of $N$ two-level atoms 
interacting with a vacuum cavity, with which we give analytic expressions for 
the value of memory effects, the cavity photon number, the degree of superradiance,
and so on. In Sec. \textrm{IV}, the influences of two types of initial states on
the memory effects as well as the superradiance are calculated and analyzed 
using the early-time results. Our work is extended to the near-Markovian regime 
and a strongly non-Markovian regime in Sec. \textrm{V}.  At last, conclusions 
and discussions are presented in Sec. \textrm{VI}.

\section{Initial-state-dependent memory effects }
In this section, we first review the measure of non-Markovianity in Ref. \cite{Hou2015},  
which quantifies the  memory effects (the past-future dependence) in a quantum process.
Using its physical interpretations, we then propose a method to quantify the memory
 effects in a quantum process conditioned on a particular initial state of the system. 
The object of study in Ref. \cite{Hou2015} is a quantum process of an open quantum system described by 
the total Hamiltonian 
\begin{eqnarray}
H=H_S+H_E+H_{SE},
\label{Eqn:TotalHmt}
\end{eqnarray}
an arbitrary system initial state $\rho_S(t_I)$ and a fixed initial 
state of the environment $\rho_E(t_I)$ (independent of the system). 
Here $t_I$ is the initial time of an evolution which is also arbitrary.  
The initial condition of an evolution is 
\begin{eqnarray}
\rho_{SE}(t_I)=\rho_{S}(t_I)\otimes\rho_E(t_I)
\label{Eqn:InitialCondition}
\end{eqnarray}
which means that the system is initially uncorrelated with the environment before the evolution. 
 In general, $\rho_E(t_I)$ is governed by \cite{Li2018}
\begin{eqnarray}
\rho_E(t_I)=\mathcal{T}e^{-i\int_{0}^{t_I}H_E(\tau)d\tau}\rho_E(0).
\label{Eqn:EnvInitialState}
\end{eqnarray}
Typically, one deals with a time-independent environment Hamiltonian $H_E$ and assumes a 
steady state of $H_E$ as the environmental initial state (e.g., a thermal state). 
Then, the initial condition of an evolution is 
\begin{eqnarray}
\rho_{SE}(t_I)=\rho_S(t_I)\otimes\rho_E
\label{Eqn:InitialConditionSteady}
\end{eqnarray}
for any $t_I$. Since $\rho_S(t_I)$ and $t_I$ is arbitrary, the measure of non-Markovianity
 (memory effects) in Ref. \cite{Hou2015} is determined by $H$ and $\rho_E(t_I)$ and the time 
 interval of interest. 

In a non-Markovian quantum process, the meaning of a dynamical map given by 
 $\rho_S(t_2)=\varepsilon(t_2,t_1)\rho_S(t_1)$ is ambiguous unless $\rho_{SE}(t_1)$, or 
 from another perspective, the initial time $t_I$ is specified ($t_I\leqslant t_1$).
For clarity, we define $T(t_2,t_1)$ as a memoryless dynamical map that transfers $\rho_S(t_1)$ to 
 $\rho_S(t_2)$,  where $t_1$ is the initial time of the evolution ($t_I=t_1$) \cite{Hou2015},
  i.e.,
 \begin{eqnarray}
  \rho_S(t_2)&=&T(t_2,t_1)\rho_S(t_1)  \nonumber
\\&=&\Tr_E[U(t_2,t_1)\rho_S(t_1)\otimes\rho_E(t_1)U(t_2,t_1)^{\dag}],
\label{Eqn:T}
\end{eqnarray}
where $\rho_E(t_1)$ is fixed, 
$\rho_{S}(t_1)$ is arbitrary and $U(t_2,t_1)=
\mathcal{T}e^{-i\int_{t_1}^{t_2}H(\tau)d\tau}$ in general. 
Particularly, when $H$ is time-independent and $\rho_E(t_I)=\rho_E$
is a steady state of $H_E$, Eq.\ (\ref{Eqn:T}) is simplified to 
 \begin{eqnarray}
\rho_S(t_2)&=&T(t_2,t_1)\rho_S(t_1) \nonumber\\
           &=&\Tr_E[e^{-iH(t_2-t_1)}\rho_S(t_1)\otimes\rho_E e^{iH(t_2-t_1)}]
\label{Eqn:homoT}
\end{eqnarray}
such that $T(t_2,t_1)=T(t_2-t_1,0)$ \cite{Chruscinski}. The dynamical map $T$ is trace-preserving
 and completely positive and called a universal  dynamical map (UDM) 
 that is independent of the state it acts upon \cite{Rivas}.  

Let $t_0\leqslant t_1\leqslant t_2$, a (time-dependent) quantum process is
 Markovian if the dynamical maps  (denoted by $\Lambda_M$) satisfy the divisibility condition 
\begin{eqnarray}
\Lambda_M(t_2,t_0)=\Lambda_M(t_2,t_1)\Lambda_M(t_1,t_0),
\label{Eqn:Mdiv}
\end{eqnarray}
where each dynamical map is uniquely defined and a UDM. Remark that the dynamical map
$\Lambda_M(t_2,t_1)$ is a UDM if and only if it is induced by 
\begin{eqnarray}
  \rho_S(t_2)&=&\Lambda_M(t_2,t_1)\rho_S(t_1) \nonumber\\
              &=&\Tr_E[U(t_2,t_1)\rho_{S}(t_1)\otimes\rho_{E}(t_1)U(t_2,t_1)^{\dag}],
\label{Eqn:UDMcondition}
\end{eqnarray}
where $\rho_{E}(t_1)$ is fixed and $\rho_{S}(t_1)$ is arbitrary \cite{Rivas}.
In an open quantum system, the condition $\rho_{SE}(t)=\rho_S(t)\otimes\rho_E(t)$
[$\rho_E(t)$  does not dependent on the system state] may not be satisfied exactly 
for $t>t_I$. Thus the dynamics of an exact open quantum system is typically 
not Markovian \cite{Rivas}. However, $\rho_{SE}(t)\approx\rho_S(t)\otimes\rho_E(t)$ 
can be a good approximation where the correlation between the system and 
the environment does not affect the system's dynamics so much \cite{Rivas}. It is observed that
the Markovian dynamical map $\Lambda_M(t_2,t_1)$ is unique and does not depend on the
initial time of an evolution, e.g., $t_I=t_1$ or $t_I<t_1$. Therefore, 
$\Lambda_M(t_2,t_1)=T(t_2,t_1)$ according to the definition of the memoryless dynamical map $T$. 
Then, the Markovian divisibility condition (\ref{Eqn:Mdiv}) can also be expressed in terms of 
the memoryless dynamical map $T$ by
\begin{eqnarray}
T(t_2,t_0)=T(t_2,t_1)T(t_1,t_0),
\label{Eqn:divT}
\end{eqnarray}
whose violation is a sign of non-Markovianity. Unlike the dynamical maps used 
in some non-Markovianity measures \cite{Rivas2010,Hou2011}, all the dynamical 
maps $T$ are completely positive. The violation of Eq.\ (\ref{Eqn:divT}) is 
manifested by the inequality 
\begin{eqnarray}
T(t_2,t_0)\neq T(t_2,t_1)T(t_1,t_0). 
\label{Eqn:ineq}
\end{eqnarray}

\begin{figure}
\includegraphics*[width=7cm]{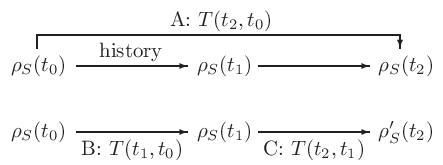}
\caption{Physical interpretation of Eq.\ (\ref{Eqn:ineq}) as memory effects.
 At the moment $t_1$, the system states in evolution A and C are the same.  
 Recall that evolution A has a history from $t_0$ to $t_1$ while evolution
 C does not have any history before $t_1$. Therefore, $\rho_S(t_2)\neq \rho'_S(t_2)$ is
 an evidence that the future (after $t_1$)  state  of the system depends on
  its history (from $t_0$ to $t_1$) in a quantum process. }                                   
\label{FIG:Maps}
\end{figure}

The physical meaning of Eq.\ (\ref{Eqn:ineq}) can be explained with Fig.\ \ref{FIG:Maps}.
Let the left-hand and the right-hand sides of Eq.\ (\ref{Eqn:ineq}) act on a system initial 
state  $\rho_S(t_0)$. On the left-hand side of Eq.\ (\ref{Eqn:ineq}), 
$\rho_S(t_0)$ is mapped to $\rho_S(t_2)$ by $T(t_2,t_0)$ in evolution A. 
On the right-hand side, $\rho_S(t_0)$ is first mapped to $\rho_S(t_1)$ 
by $T(t_1,t_0)$ in evolution B. Then, as the initial state of evolution C, 
$\rho_S(t_1)$ is mapped to $\rho'_S(t_2)$ by $T(t_2,t_1)$, which means that 
the initial condition of evolution C is  $\rho_{SE}(t_1)=\rho_{S}(t_1)\otimes\rho_{E}(t_1)$
with fixed $\rho_{E}(t_1)$ defined by Eq.\ (\ref{Eqn:EnvInitialState}) or a steady one $\rho_E$. 
 At moment $t_1$, evolution A and C have the same system state $\rho_S(t_1)$ but 
 different histories: evolution A has a history in time interval $[t_0,t_1]$, which is 
encoded in $\rho_{SE}(t_1)=U(t_1,t_0)\rho_{S}(t_0)\otimes\rho_{E}(t_0)U(t_1,t_0)^{\dag}$,
while evolution C (starting at $t_1$) has no history before $t_1$. Therefore, 
$\rho_S(t_2)\neq \rho'_S(t_2)$ is an evidence that the future state (after $t_1$) of the system 
depends on its history (in $[t_0,t_1]$) in this quantum process. This is the physical
meaning of the memory effects in  this paper and \cite{Hou2015}. Otherwise, if the divisibility 
 (\ref{Eqn:divT}) holds, the process is Markovian and $\rho_S(t_2)=\rho'_S(t_2)$ for
 any $\rho_S(t_0)$.

On the other hand, the inequality could be understood by focusing on the 
change of the environment state  between evolution B and C. At the end of evolution B, 
 $\rho_{SE}(t_1)=U(t_1,t_0)\rho_{S}(t_0)\otimes\rho_{E}(t_0)U(t_1,t_0)^{\dag}$.
Then, at the beginning of evolution C, the environment is initialized by $T(t_2,t_1)$
such that $\rho_{SE}(t_1)\rightarrow\rho_{S}(t_1)\otimes\rho_{E}(t_1)$, where $\rho_{E}(t_1)$
is independent of the system. The information of the system's history in $[t_0,t_1]$ is
erased by $T(t_2,t_1)$. In contrast, such an initialization never happens in evolution A.
Therefore, $\rho_S(t_2)\neq \rho'_S(t_2)$ signifies that the environment (as well as the 
correlations between the system and environment \cite{Rivas}) remembers the system's history and this 
memory can influence the future of the system.

In Ref. \cite{Hou2015}, we define the maximal difference of $T(t_2,t_0)$ and $T(t_2,t_1)T(t_1,t_0)$
as the non-Markovianity where $t_1$ and $t_2$ are optimized. Based on the dynamical maps $T$,
all the initial states are potentially considered, thus the measure of non-Markovianity
 in Ref. \cite{Hou2015} does not depend on the initial state of the system. In this paper, 
 our goal is to evaluate the influence of the system initial state on the memory effects 
in a quantum process. Using the physical interpretations discussed above, we define the 
value of memory effects conditioned on  $\rho_S(t_0)$ as the maximal trace 
distance between $\rho_S(t_2)$ and $\rho'_S(t_2)$:
\begin{eqnarray}
N_M[\rho_S(t_0)]=\max_{t_1,t_2} \frac{1}{2}\|\rho_S(t_2)-\rho'_S(t_2)\|.
\label{Eqn:NMI}
\end{eqnarray}
Here $||A||=\Tr(\sqrt{A^\dag A})$ is the trace norm of an operator $A$. 
Assuming a fixed $t_0$ for convenience, $N_M[\rho_S(t_0)]$ could be 
calculated by optimizing $t_1$ and $t_2$ in a time interval $[t_0,t]$
 of interest where $t_0\leqslant t_1\leqslant t_2\leqslant t$. 
For example, the time interval might be $[t_0,\infty]$ or $[t_0, t_0+\tau]$, 
 where $\tau$ is finite (as done later in this paper).  Similar to the measure in 
 Ref. \cite{Hou2015}, $0\leqslant N_M[\rho_S(t_0)] \leqslant 1$ is satisfied due
 to the properties of the trace distance. 
 
 Using Eq.\ (\ref{Eqn:NMI}), the influence of initial states on the memory effects 
 of a quantum process could be quantitatively evaluated. 
 Note that $N_M[\rho_S(t_0)]>0$ is a sufficient condition for the 
 inequality Eq.\ (\ref{Eqn:ineq}), but not a necessary one. The theoretical calculation 
 and experimental observation of Eq.\ (\ref{Eqn:NMI}) might be easier than 
 those in Ref. \cite{Hou2015} since the determination of the dynamical maps 
 $T$ is not compulsory.  Moreover, as discussed in Ref. \cite{Hou2015}, 
 $\Tr[P \rho(t_2)]\neq \Tr[P \rho'(t_2)]$ is a sufficient 
 condition for $\rho_S(t_2)\neq\rho_S'(t_2)$, where $P$ is an operator of a physical 
 quantity. Therefore, when the full information of the system state is not easily 
 obtainable, $\Delta P[\rho_S(t_0)]=|\Tr[P\rho(t_2)]-\Tr[P \rho'(t_2)]|$
 ($\max_{t_1,t_2}\Delta P[\rho_S(t_0)]$) could be used as an evidence (manifestation)
 of the memory effects for simplicity and intuitiveness both theoretically and 
 experimentally.

\section{Early-time superradiance}
\subsection{Theoretical model}
We consider a fundamental model that describes $N$ two-level atoms 
(the system, denoted by ``$S$") interacting with a cavity (the environment, denoted by ``$E$" ) 
initially in a vacuum state. The Hamiltonian is given by  
\begin{eqnarray}
H\!&=&\! H_S+H_E+H_{I} \nonumber\\
  &=&\!\omega_A\!\sum_{n=1}^{N}\!\sigma_n^+\sigma_n^-\!+\!\omega_Bb^{\dag}b\!+\!\sum_{n=1}^{N} g(\sigma_{n}^{+}b +\sigma_n^{-}b^{\dag}).
\label{Eqn:Hmt}
\end{eqnarray}
Here $H_S$ and $H_E$ represent $N$ noninteracting two-level atoms and 
a single-mode electromagnetic field in the cavity, respectively. 
$H_{I}=\sum_{n=1}^{N} g(\sigma_{n}^{+}b +\sigma_n^{-}b^{\dag})$ describes the
 interactions between the atoms and the cavity with the rotating wave approximation,
 and the coupling strength $g$ is a real constant. 
The lowering and raising operators for the $n$th atom is defined as
 $\sigma_n^-=\ket{g}_n\bra{e}_n$ and $\sigma_n^+=\ket{e}_n\bra{g}_n$. 

The initial condition of the model is assumed to be  $\rho_{S\!E}(t_0)=\rho_S(t_0)\otimes\ket{0}\bra{0}$
 where $\rho_S(t_0)$ is arbitrary and $\ket{0}$ is the vacuum state of the cavity. 
 For simplicity, we use $t_0=0$ in the remainder of this paper without loss of generality.
The density matrix of the composite system $\rho_{S\!E}$ 
is described by the master equation,
\begin{eqnarray}
\dot\rho_{S\!E}&=&-i[H_S+H_E+H_{I},\rho_{S\!E}] \nonumber\\
&&+\gamma(b\rho_{S\!E}b^{\dag}-\frac{1}{2}b^{\dag}b\rho_{S\!E}-\frac{1}{2}\rho_{S\!E}b^{\dag}b),
\label{Eqn:ME}
\end{eqnarray}
where $\gamma$ is the dissipation strength of the cavity. Assume the cavity is 
in resonance with the atoms ($\omega_A=\omega_B$), then the density matrix $\rho_{S\!E}$ 
in the interaction picture is described by 
\begin{eqnarray}
\dot\rho_{S\!E}&=&-i[\sum_{n=1}^{N}g(\sigma_{n}^{+}b +\sigma_n^{-}b^{\dag}),\rho_{S\!E}] \nonumber\\
               & &+\gamma(b\rho_{S\!E}b^{\dag}-\frac{1}{2}b^{\dag}b\rho_{S\!E}-\frac{1}{2}\rho_{S\!E}b^{\dag}b).
\label{Eqn:MEI}
\end{eqnarray} 

The superradiance phenomenon could be understood as the enhancement of collective
spontaneous radiation caused by a common environment (compared with independent environments). 
To understand the role of memory effects in the superradiance process, we are also 
interested in the dynamics where the atoms radiate independently, e.g., each
atom radiates in its own cavity. This dynamics could be described by the 
Hamiltonians $H_{I}=\sum_{n=1}^{N} g(\sigma_{n}^{+}b_n +\sigma_n^{-}b_n^{\dag})$,  
$H_E=\sum_{n=1}^{N} \omega_Bb_n^{\dag}b_n$ and the environment initial state
 $\rho_E=\Pi_{n=1}^{N}\ket{0}_n\bra{0}_n$. Under the resonance condition, 
 the master equation for the independently radiating atoms and the cavities 
 in the interaction picture is given by
 \begin{eqnarray}
\dot\rho_{S\!E}&=&-i[\sum_{n=1}^{N}g(\sigma_{n}^{+}b_n +\sigma_n^{-}b_n^{\dag}),\rho_{S\!E}] \nonumber\\
               & &+\gamma\sum_{n=1}^N(b_n\rho_{S\!E}b_n^{\dag}\!-\!\frac{1}{2}b_n^{\dag}b_n\rho_{S\!E}\!-\!\frac{1}{2}\rho_{S\!E}b_n^{\dag}b_n).
\label{Eqn:MEIInd}
\end{eqnarray}

\subsection{Early-time solution}

In this section, we focus on the early-time dynamics where $gt\ll1$ due 
to the following reasons: First, non-Markovian characters are non-negligible on such a
short timescale. Second, it stresses the influence of a particular initial state since
the state hardly changes in this time interval. Furthermore, it is helpful for understanding the 
creation mechanism of the superradiance.  In the early-time limit $gt\rightarrow0$, 
one might ignore the influence of cavity dissipation $\gamma$, providing that  $\gamma$ is not 
infinitely large. In this case,  $\rho_{S\!E}$ evolves unitarily via 
\begin{eqnarray}
\rho_{S\!E}(t)=U(t)\rho_{S}(0)\otimes\rho_{E}U^{\dag}(t),
\label{Eqn:rhoSE}
\end{eqnarray} 
where $U(t)=e^{-iH_{I}t}$ and $\rho_E=\ket{0}\bra{0}$. The influence of ignoring $\gamma$ 
on the early-time dynamics will be discussed at the end of this section. 

The reduced dynamics of the atoms could be determined by tracing out the degrees of the environment:
\begin{eqnarray}
\rho_{S}(t)&=&\Tr_E[U(t)\rho_{S}(0)\otimes\rho_EU^{\dag}(t)]\nonumber \\
               &=&\sum_{k}\bra{k}_EU(t)\rho_{S}(0)\otimes\rho_EU^{\dag}(t)\ket{k}_E,
\label{Eqn:rhoS}
\end{eqnarray} 
where $\ket{k}_E$ are a set of basis in $\mathcal{H}_E$. 
With the help of the Baker-Campbell-Hausdorff formula 
\begin{eqnarray}
e^{\alpha A}Be^{-\alpha A}=B+\alpha [A,B]+\frac{\alpha^2 }{2!}[A,[A,B]]+\cdots,
\label{Eqn:BCH}
\end{eqnarray} 
after the replacement $A=H_I$, $B=\rho_{S}(0)\otimes\rho_E$, and $\alpha=-it$, we have 
\begin{eqnarray}
\rho_{S}(t)&\approx&\sum_{k}\bra{k}(\rho_S(0)\otimes\rho_E-it[H_I,\rho_S(0)\otimes\rho_E]\nonumber\\
&&-\frac{1}{2}t^2[H_I,[H_I, \rho_S(0)\otimes\rho_E]]\ket{k}
\label{Eqn:2Order}
\end{eqnarray} 
where the terms with higher orders of $gt$ have been omitted due to the early-time limit ($gt\rightarrow0$). 
Using the number state basis of the cavity $\ket{k}=\ket{0},\ket{1},\cdots$ and
 the assumption $\rho_E=\ket{0}\bra{0}$, it is straightforward to derive the solution
  of the system evolution  in the early-time limit $gt\rightarrow0$,
\begin{eqnarray}
\rho_{S}(t)=\rho_S(0)+(gt)^2\mathcal{L}_{\sigma^-}[\rho_S(0)]. 
\label{Eqn:AtmDyn}
\end{eqnarray} 
Here $\sigma^-=\sum_n\sigma_n^-$ is the collective lowering operator and 
 $\mathcal{L}_{K}(\rho) = K\rho K^+-\frac{1}{2}K^\dag K\rho-\frac{1}{2}\rho K^\dag K$ is the Lindblad superoperator.
 $b^{\dag}\ket{0}=\ket{1}$ and $\bra{0}b=\bra{1}$ have been used to obtain Eq.\ (\ref{Eqn:AtmDyn}).  
Notice that the partial trace over  the second term of Eq.\ (\ref{Eqn:2Order}) is zero, 
thus the change of $\rho_S(0)$ is quadratic in $t$ in 
the early-time limit. 
Using similar procedures done above, we obtain the early-time solution 
for independently radiating atoms:
 \begin{eqnarray}
\rho_{S}(t)=\rho_S(0)+(gt)^2\sum_{n=1}^{N}\mathcal{L}_{\sigma_n^-}[\rho_S(0)]. 
\label{Eqn:AtmDynInd}
\end{eqnarray}

\subsection{Memory effects}
With the above results, we now calculate the initial-state-dependent memory effects defined 
in Sec.\textrm{II}. We consider the dynamics in a short time interval $[0,t]$
where $t_0=0\leqslant t_1\leqslant t_2\leqslant t$.   In ``evolution A'' (with a history from $t_0$ to $t_1$)
as mentioned before, the initial state is  $\rho_S(0)$ and the final state $\rho_{S}(t_2)$ is given by 
\begin{eqnarray}
\rho_{S}(t_2)&=&\rho_{S}(0)+[g(\tau_{10}+\tau_{21})]^2\mathcal{L}_{\sigma^-}[\rho_{S}(0)].
\label{Eqn:rhot2}
\end{eqnarray} 
Here $\tau_{10}=t_1-t_0$ and $\tau_{21}=t_2-t_1$ are used for convenience. 
The initial state of  ``evolution C'' at $t_1$  is 
\begin{eqnarray}
\rho_{S}(t_1)&=&\rho_S(0)+(g\tau_{10})^2\mathcal{L}_{\sigma^-}[\rho_S(0)].
\label{Eqn:rhot1}
\end{eqnarray} 
In ``evolution C'' (without any history before $t_1$), the final state  $\rho'_{S}(t_2)$ is given by 
\begin{eqnarray}
\label{Eqn:rhot2p}
\rho'_{S}(t_2)&=&\rho_{S}(t_1)+(g\tau_{21})^2\mathcal{L}_{\sigma^-}[\rho_{S}(t_1)].
\end{eqnarray} 
Substituting Eq.\ (\ref{Eqn:rhot1}) into Eq.\ (\ref{Eqn:rhot2p}), we have 
\begin{eqnarray}
\rho'_{S}(t_2)&=&\rho_S(0)+(g\tau_{10})^2\mathcal{L}_{\sigma^-}[\rho_S(0)]\nonumber \\
                 &&+(g\tau_{21})^2\mathcal{L}_{\sigma^-}\{\rho_S(0)+(g\tau_{10})^2\mathcal{L}_{\sigma^-}[\rho_S(0)]\}\nonumber\\
                 &=&\rho_S(0)+(g\tau_{10})^2\mathcal{L}_{\sigma^-}[\rho_S(0)]\nonumber\\
                 &&+(g\tau_{21})^2\mathcal{L}_{\sigma^-}[\rho_S(0)]
\label{Eqn:rhot2p2}
\end{eqnarray} 
where the high-order term $g^4\tau_{10}^2\tau_{21}^2\mathcal{L}_{\sigma^-}\{\mathcal{L}_{\sigma^-}[\rho_S(0)]\}$ 
has been omitted due to the early-time limit. The memory effects are manifested by the 
difference between  Eq.\ (\ref{Eqn:rhot2}) and Eq.\ (\ref{Eqn:rhot2p2}):
\begin{eqnarray}
\rho_{S}(t_2)\!-\!\rho'_{S}(t_2)\!&=&\!g^2[(\tau_{10}+\tau_{21})^2\!-\!\tau_{10}^2\!-\!\tau_{21}^2]\mathcal{L}_{\sigma^-}[\rho_S(0)]\nonumber\\
            &=&\!2g^2\tau_{10}\tau_{21}\mathcal{L}_{\sigma^-}[\rho_S(0)].
\label{Eqn:Difrhos}
\end{eqnarray}   
According to Eq.\ (\ref{Eqn:NMI}), the value of the memory effects for the initial state $\rho_S(0)$ is  
\begin{eqnarray}
N_M[\rho_S(0)]&=& \max_{\tau_{10},\tau_{21}}\frac{1}{2}\|\rho_{S}(t_2)-\rho'_{S}(t_2)\|\nonumber\\
              &=& g^2\|\mathcal{L}_{\sigma^-}[\rho_S(0)]\|\max_{\tau_{10},\tau_{21}}(\tau_{10}\tau_{21}).
\label{Eqn:NMSR}
\end{eqnarray}   
Since $g^2\|\mathcal{L}_{\sigma^-}[\rho_S(0)]\|$ is a constant for a given $\rho_S(0)$, 
Eq.\ (\ref{Eqn:NMSR}) can be calculated simply by maximizing $\tau_{10}\tau_{21}$ with 
the constraint $0\leqslant\tau_{10}+\tau_{21}=\tau_{20}\leqslant t$. 
Let $\tau_{10}+\tau_{21}=t'$, then $\tau_{10}\tau_{21}=-(\tau_{10}-\frac{t'}{2})^2+\frac{t'^2}{4}\leqslant \frac{t'^2}{4}$.
Therefore, the maximum of $\tau_{10}\tau_{21}$ is $\frac{t'^2}{4}$ when $\tau_{10}=\tau_{21}=\frac{t'}{2}$. 
It is easy to see that the maximum of $\tau_{10}\tau_{21}$ in the time interval  $[0,t]$ 
happens when $t'=t$. Eventually,  $N_M[\rho(t_0)]$ 
in the early-time limit is 
\begin{eqnarray}
N_M[\rho_S(0)]&=&g^2\|\mathcal{L}_{\sigma^-}[\rho_S(0)]\|\max_{\tau_{10},\tau_{21}}(\tau_{10}\tau_{21})\nonumber\\
              &=&g^2\|\mathcal{L}_{\sigma^-}[\rho_S(0)]\| (\frac{t}{2})^2   \nonumber\\
              &=&\frac{1}{4}(gt)^2 \|\mathcal{L}_{\sigma^-}[\rho_S(0)]\|.
\label{Eqn:NMopt}
\end{eqnarray}  
Equation (\ref{Eqn:NMopt}) demonstrates that  in a short time interval $[0,t]$,
 $N_M[\rho(0)]$ grows quadratically with $t$. To focus on the influence 
of initial states (rather than the time $t$  or the atom-cavity coupling $g$), 
it is convenient to discuss the normalized value of memory effects
\begin{eqnarray}
\frac{N_M[\rho_S(0)]}{(gt)^2}=\frac{1}{4}\|\mathcal{L}_{\sigma^-}[\rho_S(0)]\|. 
\label{Eqn:NMNorm}
\end{eqnarray}
It represents the strength of memory effects in the early-time limit as a function 
of only $\rho_S(0)$. Notice that $\frac{N_M[\rho_S(0)]}{(gt)^2}$ is not bounded as $N_M[\rho_S(0)]$. 
In principle, there is  $0\leqslant\frac{N_M[\rho_S(0)]}{(gt)^2}<\infty$.
Similarly, the normalized memory effects for independent cavities (denoted by $N_M^{ind}$) 
can be derived from Eq.\ (\ref{Eqn:AtmDynInd}). The result can be expressed simply by 
replacing   $\mathcal{L}_{\sigma^-}[\rho_S(0)]$ in Eq.\ (\ref{Eqn:NMNorm}) by 
$\sum_{n=1}^N\mathcal{L}_{\sigma_n^-}[\rho_S(0)]$, i.e.,
\begin{eqnarray}
\frac{N_M^{\textrm{ind}}[\rho_S(0)]}{(gt)^2}=\frac{1}{4}\|\sum_{n=1}^N\mathcal{L}_{\sigma_n^-}[\rho_S(0)]\|. 
\label{Eqn:NMNormInd}
\end{eqnarray}

\subsection{Cavity photon number}
In a non-Markovian proceess, the relation $\rho_{SE}(t)\approx\rho_S(t)\otimes\rho_E$ 
does not hold in general. Particularly, for the superradiance problem, the change of photon 
number in the environment (the cavity) might be an important source of the memory effects
so that $\rho_{SE}(t)\neq\rho_S(t)\otimes\rho_E$.  Thus it is desirable to know the cavity
photon number (denoted by $N_P$ in this paper) in the early-time limit 
in order to  understand the physics of the memory effects. The total excitation number
of our model represented by $\hat{N}_{ex}= b^\dag b+\sum_n\sigma_n^+\sigma_n^-$ is conserved
since $[\hat{N}_{ex},H]=0$. Besides, there are no photon in the cavity initially. Therefore,
the cavity photon number is equal to the loss of excitations of the atoms that represents 
the emission intensity of superradiance (in the early-time limit).  

According to Eq.\ (\ref{Eqn:AtmDyn}), the cavity photon number at $t$ for $\rho_S(0)$ is given by,
 \begin{eqnarray}
N_P[\rho_S(0)]&=&\Tr[\sum_n\sigma_n^+\sigma_n^-\rho_S(0)]-\Tr[\sum_n\sigma_n^+\sigma_n^-\rho_S(t)]\nonumber\\
      &=&\Tr[\sum_n\sigma_n^+\sigma_n^-\rho_S(0)]-\Tr[\sum_n\sigma_n^+\sigma_n^-\rho_S(0)]\nonumber\\
      & &-\Tr\{\sum_n\sigma_n^+\sigma_n^-(gt)^2\mathcal{L}_{\sigma^-}[\rho_S(0)]\}\nonumber\\
      &=&(gt)^2\Tr\{-\sum_n\sigma_n^+\sigma_n^-\mathcal{L}_{\sigma^-}[\rho_S(0)]\}.
\label{Eqn:Np}
\end{eqnarray} 
Using $[\sigma^+\sigma^-,\sum_n\sigma_n^+\sigma_n^-]=0$ and 
 $[\sum_n\sigma_n^+\sigma_n^-,\sigma^+]=\sigma^+$, Eq.\ (\ref{Eqn:Np}) can be simplified to 
 a more concise form
  \begin{eqnarray}
N_P[\rho_S(0)]=(gt)^2\Tr[\sigma^+\sigma^-\rho_S(0)].
\label{Eqn:NpSmp}
\end{eqnarray} 
It is observed that in the early-time limit, the cavity number increases quadratically with time. 
As mentioned in the last section, it is convenient to discuss the normalized 
cavity photon number
\begin{eqnarray}
\frac{N_P[\rho_S(0)]}{(gt)^2}=\Tr[\sigma^+\sigma^-\rho_S(0)]
\label{Eqn:NpNorm}
\end{eqnarray} 
such that it does not depend on the evolution time $t$ and atom-cavity coupling $g$. 
Equation (\ref{Eqn:NpNorm}) represents the emission intensity in the early-time limit as a
function of only the initial state. When each atom radiates independently, e.g., 
each atom radiates in its own cavity, the normalized cavity photon number for
 the $n$th atom is reduced to 
\begin{eqnarray}
\frac{N_P^{(n)}[\rho_{S_n}(0)]}{(gt)^2}=\Tr[\sigma_n^+\sigma_n^-\rho_{S_n}(0)],
\label{Eqn:NpNormInd}
\end{eqnarray}
where $\rho_{S_n}(0)$ is the reduced density matrix of the $n$th atom. Then
the degree of superradiant in the early-time regime might be measured by the ratio \cite{Rehler,Kim}
\begin{eqnarray}
S[\rho_{S}(0)]&=&\frac{N_P[\rho_S(0)]}{\sum_nN_P^{(n)}[\rho_S(0)]}\nonumber\\
              &=&\frac{\Tr[\sigma^+\sigma^-\rho_{S}(0)]}{\sum_n\Tr[\sigma_n^+\sigma_n^-\rho_{S_n}(0)]}
\label{Eqn:DgrSR}
\end{eqnarray}
for nonzero denominator. If $S[\rho_{S}(0)]$ is greater (less) than one,
 the state $\rho_{S}(0)$ is superradiant (subradiant). 
\subsection{Other manifestations of memory effects}
As discussed in Sec. \textrm{II}, the difference between $\Tr[P \rho(t_2)]$ and $\Tr[P \rho'(t_2)]$ 
can be used as an evidence or a manifestation of the memory effects. For the superradiance problem, 
the atom excitation number operator $P=\sum_n\sigma_n^+\sigma_n^-$ serves as a nature choice. 
Following similar treatments in Eqs.\ (\ref{Eqn:Difrhos})-(\ref{Eqn:NMopt}),
it is seen that the maximum of the atom excitation difference $\Delta N_{ex}^{\textrm{atom}}[\rho_S(0)]=|\Tr[\sum_n\sigma_n^+\sigma_n^-\rho_S(t_2)]-\Tr[\sum_n\sigma_n^+\sigma_n^-\rho'_S(t_2)]|$ 
in the time interval $[0,t]$ also happens at $\tau_{10}=\tau_{21}=\frac{t}{2}$ and 
\begin{eqnarray}
& & \max_{\tau_{10},\tau_{21}}\Delta N_{ex}^{\textrm{atom}}[\rho_S(0)]\nonumber\\
&=&|2g^2\Tr\{\sum_n\sigma_n^+\sigma_n^-(\frac{t}{2})^2\mathcal{L}_{\sigma^-}[\rho_S(0)]\}|. 
\label{Eqn:Manifestation}
\end{eqnarray}
Using the results given by Eq.\ (\ref{Eqn:Np}) and (\ref{Eqn:NpSmp}), the manifestation 
can be further simplified by
\begin{eqnarray}
\max_{\tau_{10},\tau_{21}}\Delta N_{ex}^{\textrm{atom}}[\rho_S(0)]&=&\frac{(gt)^2}{2} \Tr[\sigma^+\sigma^-\rho_S(0)]\nonumber\\
&=&\frac{1}{2} N_P[\rho_S(0)].
\label{Eqn:ManifestationSmp}
\end{eqnarray}
It is seen that the maximal difference of the atom excitation numbers in $[0,t]$ is exactly
half the cavity photon number at $t$ in the early-time limit. Therefore, the nonzero cavity photon number 
itself is also a manifestation of memory effects.

 \subsection{Early-time regime}

Mathematically, Eq.\ (\ref{Eqn:AtmDyn}) holds true as $gt\rightarrow 0$.
One may wonder the validity of the early-time solution Eq.\ (\ref{Eqn:AtmDyn})
in a longer time interval and how the cavity dissipation deteriorates the validity. 
In this section, we discuss this problem by comparing the systems dynamics 
by Eq.\ (\ref{Eqn:AtmDyn}) with that by numerically solving the full dynamics in 
 $\mathcal{H}_S\otimes\mathcal{H}_E$ with Eq.\ (\ref{Eqn:MEI}) and tracing out 
the environment. A dynamics could be represented by the dynamical maps $T(t,0)$
 that turns all the possible initial states at $0$ to their final states at $t$. Furthermore, 
the difference of two dynamical maps can be evaluated  through their
 Choi-Jami\'{o}{\l}kowski matrices \cite{Choi,Jamiolkowski}.
Here we evaluate the error of Eq.\ (\ref{Eqn:AtmDyn})
 at instant $t$ by the trace distance of $\rho_{T(t,0)}^{\textrm{quad}}$ and $\rho_{T(t,0)}^{\textrm{exact}}$, i.e.,
\begin{eqnarray}
Error=\frac{1}{2}\|\rho_{T(t,0)}^{\textrm{quad}}-\rho_{T(t,0)}^{\textrm{exact}}\|
\label{Eqn:Error}
\end{eqnarray} 
where $\rho_{T(t,0)}^{\textrm{quad}}$ and  $\rho_{T(t,0)}^{\textrm{exact}}$ are the 
Choi-Jami\'{o}{\l}kowski matrices calculated by  Eq.\ (\ref{Eqn:AtmDyn}) 
and  (\ref{Eqn:MEI}), respectively. The  Choi-Jami\'{o}{\l}kowski 
matrix is defined as 
$\rho_T(t,0)=T(t,0)\otimes\mathbb{I}(\ket{\Psi}\bra{\Psi})$ with 
$\ket{\Psi}=\frac{1}{\sqrt{2^N}}\sum_{i=1}^{2^N}\ket{i}\ket{i}$ 
a maximally entangled state of the $N$-atom system and an ancillary 
system of the same dimension.

  \begin{figure}
\includegraphics*[width=9cm]{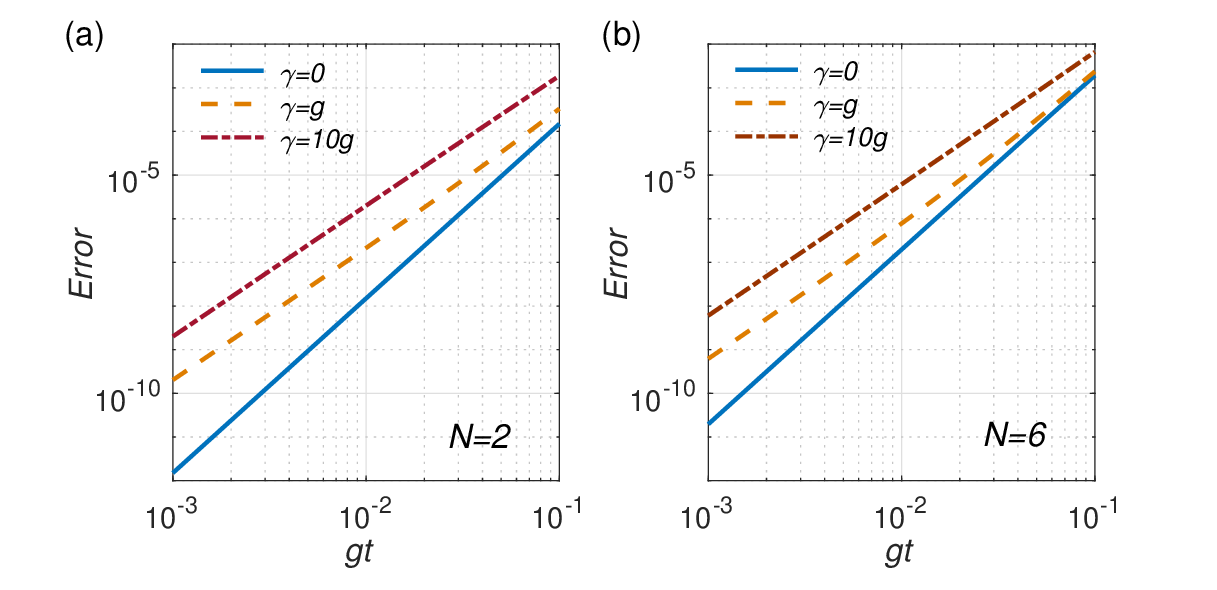}
\caption{The error of Eq.\ (\ref{Eqn:AtmDyn}) as a function of $gt$ for (a) 2 atoms and (b) 6 atoms
with different strengths of cavity dissipation. The error represents the difference of the dynamical maps
 $T(t,0)$ corresponding to Eq.\ (\ref{Eqn:AtmDyn}) (early-time assumption)  and Eq.\ (\ref{Eqn:MEI}) 
 (exact). The error vanishes as $gt\rightarrow0$ and increases with the 
 cavity dissipation $\gamma$ or the atom number $N$. }                                   
\label{FIG:Error}
\end{figure} 

\begin{figure}
\includegraphics*[width=9cm]{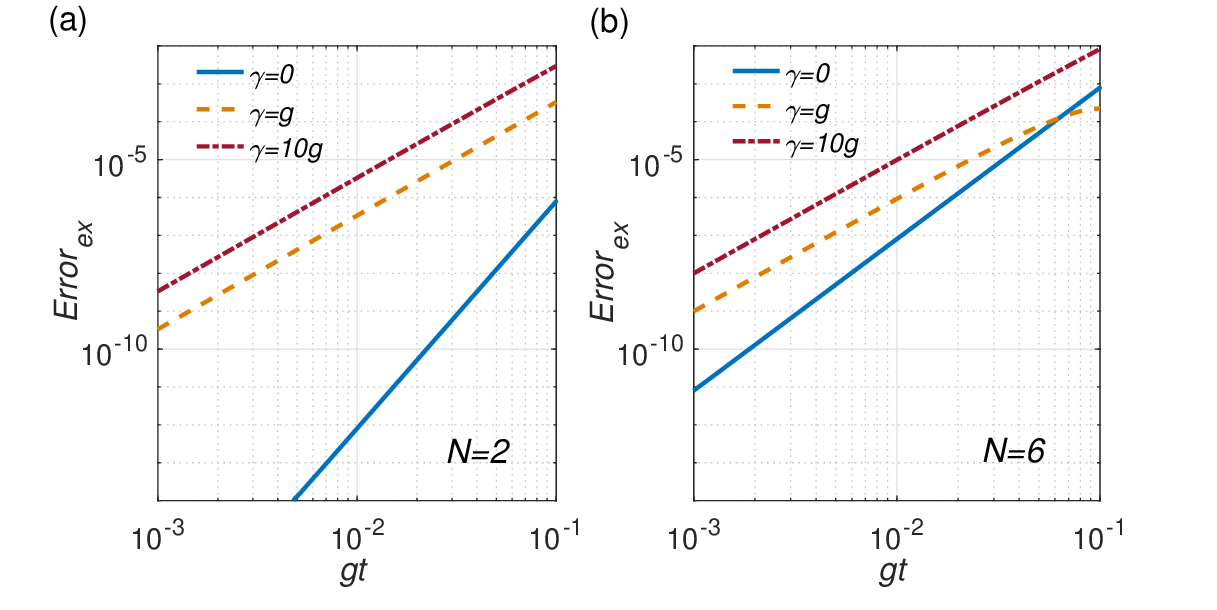}
\caption{
The error of Eq.\ (\ref{Eqn:AtmDyn}) in terms of the atom excitation number 
difference  for (a) 2 and (b) 6 fully excited atoms.  Other parameters are 
the same with those in Fig.\ \ref{FIG:Error}.}                                
\label{FIG:NError}
\end{figure}

We numerically calculated the error of Eq.\ (\ref{Eqn:AtmDyn}) for 
$\gamma/g=0,1,10$ and $N=2,6$. The results are illustrated in Fig.\ \ref{FIG:Error} 
in logarithmic scale. It is seen that the order of magnitude of error 
decreases linearly  with that of $t$. Although the cavity dissipation and the
increasing of atom number can increase the error, there exists a time interval 
where Eq.\ (\ref{Eqn:AtmDyn}) is a good approximation as long as  $\gamma$ and $N$
are finite.  For example, in a time interval $[0, 0.01/g]$,  the error is less 
than $10^{-5}$ even when $\gamma=10g$ for both $N=2$ and $N=6$,  implying that 
 Eq.\ (\ref{Eqn:AtmDyn}) is a good approximation in this scenario. 
 The above discussion implies that the early-time limit $gt\rightarrow0$ could
be relaxed to an early-time regime, i.e., a time interval $[0,\tau]$ 
where the decay is quadratic. Here $\tau$ is linked to the Zeno time \cite{Facchi,Schulman,Crespi,Wu2017}.  
We also illustrate the difference of the atom excitation numbers 
from Eqs.\ (\ref{Eqn:AtmDyn}) and  (\ref{Eqn:MEI}), i.e., $Error_{ex}=|\langle\sum_n\sigma_n^+\sigma_n^-\rangle_{quad}-
\langle\sum_n\sigma_n^+\sigma_n^-\rangle_{exact}|$, in Fig.\ \ref{FIG:NError}
as another manifestation of the error. The initial states in Figs.\ \ref{FIG:NError}(a) and \ref{FIG:NError}(b)
are 2 and 6 fully excited atoms, respectively. The result is similar to 
that in Fig.\ \ref{FIG:Error} but easier to be verified experimentally.

\section{Influence of initial states}
In this section, the results Eqs.\ (\ref{Eqn:NMNorm}),  (\ref{Eqn:NMNormInd}),
 (\ref{Eqn:NpNorm}) and  (\ref{Eqn:DgrSR}) are applied to several 
types of $N$-atom initial states.  Our first aim is to study the influence of 
initial states on the memory effects. Another aim is to reveal the role of 
memory effect in creating superradiance. Conversely, it also helps to 
understand the physical source of memory effects in our model.
\subsection{Dicke states}
The Dicke states, written as $\ket{JM}$, are extensively studied in the field of superradiance. 
It is defined as the common eigenstate of the pseudospin operators 
$D^2=\frac{1}{2}(D^+D^-+D^-D^+)+D_z^2$ and  $D_z$ with $J=N/2$. The eigenvalues 
are given by
\begin{eqnarray}
D^2\ket{JM}&=&J(J+1)\ket{JM}\\
D_z\ket{JM}&=&M\ket{JM} \qquad (M=-J,\cdots,J)
\label{Eqn:EigenDicke}
\end{eqnarray}
where the operators are defined by  $D^+=\sum_n\sigma_n^+$,  $D^-=\sum_n\sigma_n^-$, and  
 $D_z=\frac{1}{2}\sum_n(\ket{e}_n\bra{e}_n-\ket{g}_n\bra{g}_n)$.
A Dicke state is symmetrical and invariant by atom permutation
with $N_e=J+M$ excited atoms and $N_g=J-M$ ground-state atoms. It can be constructed 
by the following formula  \cite{Gross}:
\begin{eqnarray}
\ket{JM}=\sqrt{\frac{(J+M)!}{N!(J-M)!}}(\sum_n\sigma_n^-)^{J-M}\ket{e,e,\cdots,e}.
\label{Eqn:JMstate}
\end{eqnarray} 
For example, for 3 two-level atoms, there is 
$\ket{3/2,-1/2}=\frac{1}{\sqrt{3}}(\ket{egg}+\ket{geg}+\ket{gge})$.

A well-known result of the free-space spontaneous emission rate for
two-level atoms is $W_N=\Gamma\langle \sigma^+\sigma^-\rangle$, 
where $\Gamma$ is the natural linewidth of a single atom.
For the Dicke states, there is
\begin{eqnarray}
W_N&=&\Gamma\bra{JM}\sigma^+\sigma^-\ket{JM}\nonumber\\
   &=&\Gamma(J+M)(J-M+1).
\label{Eqn:JMEmissionRateMarkov}
\end{eqnarray} 
When $M=J$ (fully excited), $W_N=N\Gamma$. Then the emission rate is proportional to $N$, or say, 
the $N$-atom emission rate is equal to the summation of those from independent atoms.
Thus there is no superradiance. When $M=0$ (half excited),  $W_N=\Gamma\frac{N}{2}(\frac{N}{2}+1)$, 
the emission rate increase with $N^2$, or say, the $N$-atom emission rate is greater than 
the summation of those from independent atoms ($\frac{N}{2}\Gamma$). The superradiance 
with the most intensive emission happens in this case. Note that the results holds under
the Born-Markov approximation and on a coarse-grained timescale \cite{Gross,Agarwal1970,Bonifacio.I,Bonifacio.II,Carmichael}.

\begin{figure}
\includegraphics*[width=7.5cm]{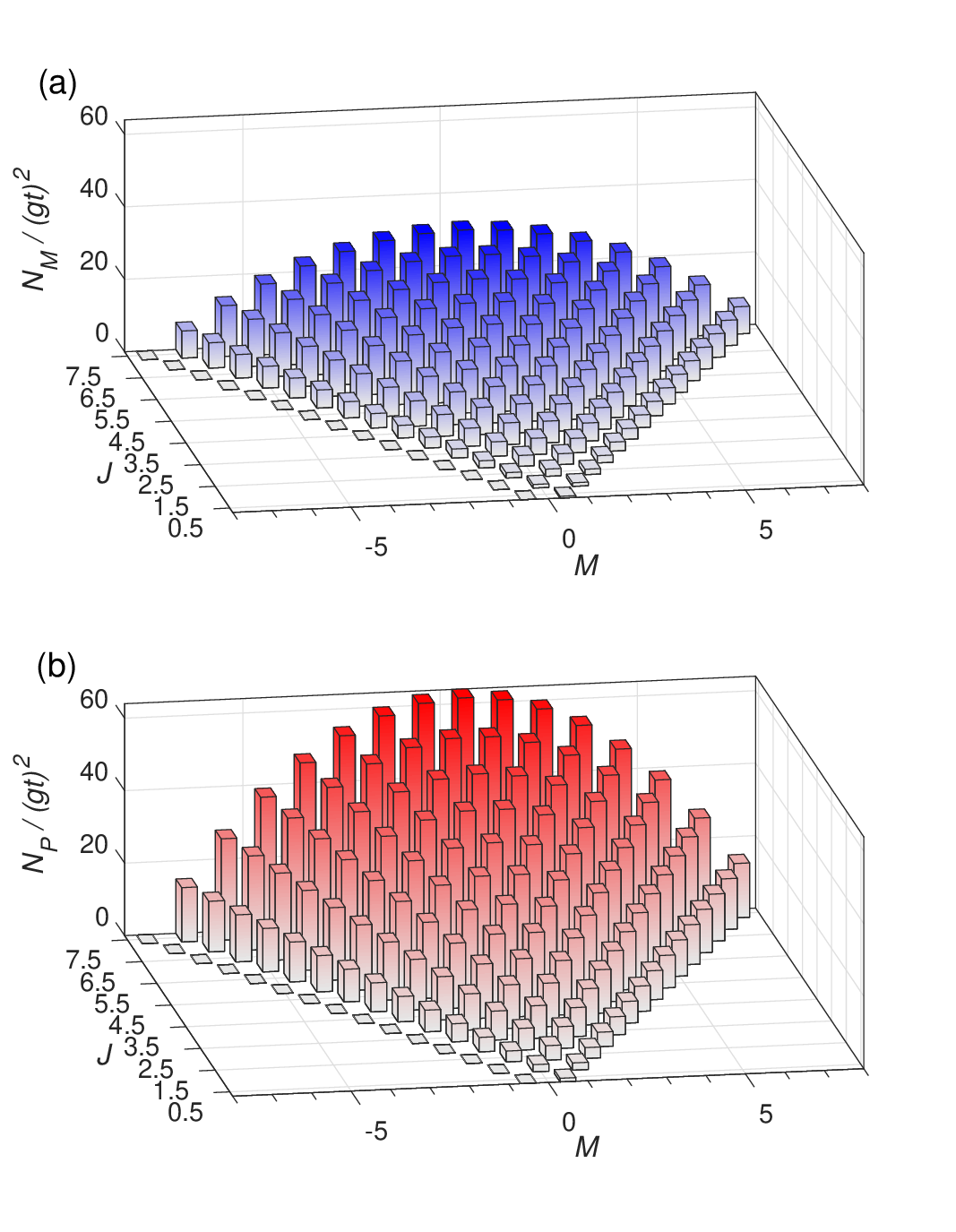}
\caption{(a) Normalized memory effects $N_M$ and (b) normalized cavity photon number $N_P$ 
 in the early-time regime  for different Dicke states $\ket{JM}$. 
 The number of atoms $N$ varies from 1 to 15, correspondingly, $J=N/2=1/2,3/2,\cdots,15/2$.
 The results demonstrate that if the atoms are initially in $\ket{JM}$, 
 the normalized value of memory effects is half the normalized cavity photon number 
 in the early-time regime. The strongest memory effects as well as 
photon emission happens when $M$ is 0 or next to 0.}
\label{FIG:NmNpJM}
\end{figure}

In the non-Markovian early-time regime, we calculated the normalized value of memory effects
and cavity photon number with Eqs.\ (\ref{Eqn:NMNorm}) and (\ref{Eqn:NpNorm}) for the Dicke states of different $N$. 
The results are shown in Fig.\ \ref{FIG:NmNpJM} where $N=1,2,\cdots,15$ ($J=N/2$) and $M=-J,-J+1,\cdots,J$.
It is observed that when the number of atoms $N$ is fixed, the strongest memory effects
as well as photon emission happens when $M$ is 0 or  next to 0, which is in 
agreement with Eq.\ (\ref{Eqn:JMEmissionRateMarkov}).  Interestingly, we find that
the value of memory effects for the Dicke states can be fully determined by the
 cavity photon number in the early-time regime:
\begin{eqnarray}
N_M(\ket{JM})&=&\frac{1}{2}N_P(\ket{JM}).
\label{Eqn:NMvsNp}
\end{eqnarray} 
Considering the result in Eq.\ (\ref{Eqn:ManifestationSmp}), there is also 
\begin{eqnarray}
N_M(\ket{JM})=\max_{\tau_{10},\tau_{21}}\Delta N_{ex}^{\textrm{atom}}[\ket{JM}].
\label{Eqn:NMvsDifNex}
\end{eqnarray}
Thus the value of memory effects can be directly measured through the atom excitation number difference 
or the cavity photon number. The relation Eq.\ (\ref{Eqn:NMvsNp}) is proved in the following.
 Using the property $\sigma^\pm\ket{JM}=\sqrt{J(J+1)-M(M\pm1)}\ket{J(M\pm1)}$,
it is straightforward to see that
\begin{eqnarray}
\frac{N_M(\ket{JM})}{(gt)^2}\!&=&\!\frac{1}{4}\|\mathcal{L}_{\sigma^-}(\ket{JM})\|\nonumber\\
                              &=&\!\frac{1}{4}\|\sigma^-\ket{JM}\bra{JM}\sigma^+\!-\!\frac{1}{2}\sigma^+\sigma^-\ket{JM}\bra{JM}\nonumber\\
                              & &\!-\frac{1}{2}\ket{JM}\bra{JM}\sigma^+\sigma^-\|\nonumber\\
                              &=&\!\frac{1}{4}(J\!+\!M)(J\!-\!M\!+\!1)\big\| \ket{JM'}\bra{JM'}\nonumber\\
                              & &-\ket{JM}\bra{JM} \big\|
\label{Eqn:NMJMprov}
\end{eqnarray}
where $M'=M-1$.  Remind that the trace norm could be calculated by 
$\|A\|=\Tr\sqrt{A^\dag A}=\sum_m|\lambda_m|$, where $\lambda_m$ 
is the eigenvalue of operator $A$. Meanwhile, the eigenvalues of 
$\ket{JM'}\bra{JM'}-\ket{JM}\bra{JM}$ is $1$ and $-1$ whose absolute 
values sum up to 2.  Therefore, 
\begin{eqnarray}
\frac{N_M(\ket{JM})}{(gt)^2}=\frac{1}{2}(J+M)(J-M+1).
\label{Eqn:NMJM}
\end{eqnarray} 
Considering that the normalized cavity photon number for the Dicke states is
 \begin{eqnarray}
\frac{N_P(\ket{JM})}{(gt)^2}&=&\Tr(\sigma^+\sigma^-\ket{JM}\bra{JM})\nonumber\\
                            &=&\Tr[(J+M)(J-M+1)\ket{JM}\bra{JM}]\nonumber\\
                            &=&(J+M)(J-M+1),
\label{Eqn:NpJM}
\end{eqnarray} 
Eq.\ (\ref{Eqn:NMvsNp}) is proved. The relation Eq.\ (\ref{Eqn:NMvsNp})  demonstrates that 
the (change of) the cavity photon number is a fundamental reason of the memory effects 
in the early-time regime for the Dicke states. 

 \begin{figure}
\includegraphics*[width=7.5cm]{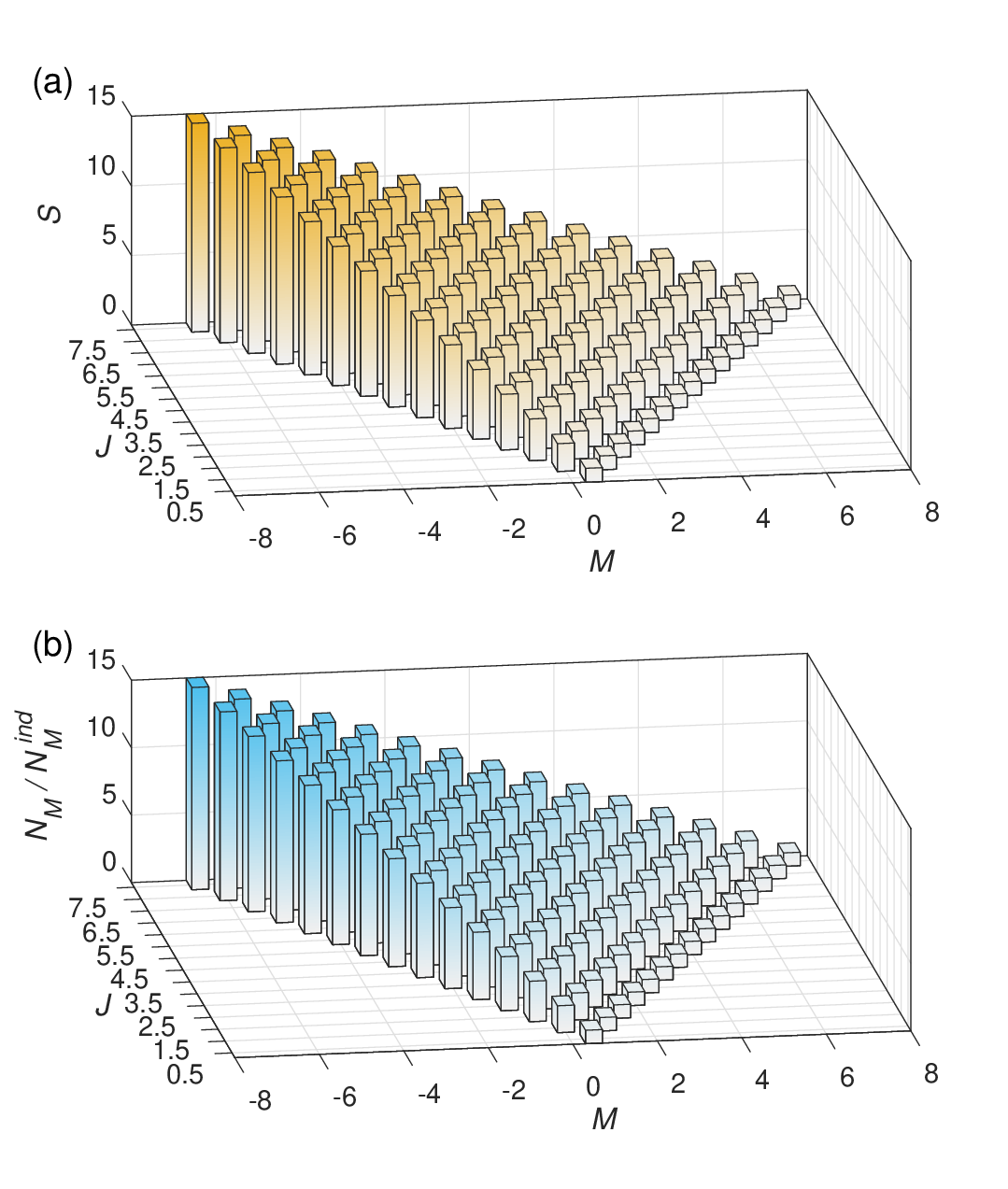}
\caption{
(a) The degree of superradiance $S$ and (b) the degree of memory-effect-enhancement  
$N_M/N_M^{\textrm{ind}}$ (from independent radiation to collective radiation) in the early-time regime 
for different Dicke states. The states are the same with those in Fig.\ \ref{FIG:NmNpJM} except 
that $S$ and $N_M/N_M^{\textrm{ind}}$ for $M=-J$ are undefined and not shown. For a given $J$, the highest degree of superradiance happens when $M=-J+1$ (single excitation). }
\label{FIG:DegreeNMComInd}
\end{figure}

We now discuss the degree of superadiance for the Dicke states in the early-time regime.
The normalized cavity photon number for one independent atom in its excited state is 
given by $N_P(\ket{e})/(gt)^2=N_P(\ket{\frac{1}{2},\frac{1}{2}})/(gt)^2=1$ by Eq.\ (\ref{Eqn:NpNormInd}).
Using Eq.\ (\ref{Eqn:NpJM}), the degree of superradiance for the Dicke states 
could be calculated by 
\begin{eqnarray}
S(\ket{JM})&=&\frac{N_P(\ket{JM})}{\sum_{n=1}^{J+M}N_P(\ket{e})} \nonumber\\
           &=& \frac{(J+M)(J-M+1)}{J+M}\nonumber\\
           &=& J-M+1
\label{Eqn:DgrSRDicke}
\end{eqnarray}
for $J+M>0$ as shown in Fig.\ \ref{FIG:DegreeNMComInd}(a). 
Here the denominator represents the contribution of the emission from the $J+M$
independently excited atoms in $\ket{JM}$. Note that the physical meaning of the denominator 
in Eq.\ (\ref{Eqn:DgrSRDicke}) is different from that in Eq.\ (\ref{Eqn:DgrSR}), 
but their values are the same.  In Fig.\ \ref{FIG:DegreeNMComInd}(a), $S$ for $\ket{J,-J}$ (no atom excited) are undefined  and not shown. Another group of Dicke states without superradiance are $\ket{J,J}$  (fully excited) corresponding to the right row in Fig.\ \ref{FIG:DegreeNMComInd}(a). 
Except for the two groups, other Dicke states are 
superradiant which could be examined by $S(\ket{JM})>1$ with Eq.\ (\ref{Eqn:DgrSRDicke}). 
For example, for $\ket{J,-J+1}$  (one atom excited)
represented by the left row in Fig.\ \ref{FIG:DegreeNMComInd}(a), there is $S(\ket{J,-J+1})=N$.
The single-photon superradiance is called ``the greatest radiation anomaly'' by Dicke. 
For the states in the right row in Fig.\ \ref{FIG:DegreeNMComInd}(a), 
there is $S(\ket{J,J-1})=2$ for $N>1$, showing a weaker degree of superradiance.

The degree of superradiance represents the magnification  
of the radiation intensity caused by a common environment (compared with $N$ independent environments). 
One interesting question is whether the memory effects are simultaneously enhanced by a 
common environment. To uncover the role of memory effects in the superradiance process, 
we next examine the degree of memory-effect-enhancement, i.e., $N_M(\ket{JM})/N_M^{\textrm{ind}}(\ket{JM})$, 
where $N_M^{\textrm{ind}}(\ket{JM})$ is the value of memory effects with $N$ independent cavities
 calculated by Eq.\ (\ref{Eqn:NMNormInd}). We find by mathematical induction that the value 
 of memory effects for independently radiating atoms in a Dicke state is 
 $N_M^{\textrm{ind}}(\ket{JM})=\frac{1}{2}(gt)^2(J+M)$. Therefore, the degree of
  memory-effect-enhancement in the early-time regime is 
\begin{eqnarray}
\frac{N_M(\ket{JM})}{N_M^{\textrm{ind}}(\ket{JM})}=J-M+1=S(\ket{JM})
\label{Eqn:Dmee}
\end{eqnarray} 
as shown in Fig.\ \ref{FIG:DegreeNMComInd}(b). The enhancement of radiation intensity 
(caused by a common environment) is equal to the enhancement of memory effects for 
the Dicke states, implying that the memory effects of a common environment 
contribute to the establishment of superradiance.

\subsection{Factorized identical states}
In this section, we investigate $N$-atom initial states in a factorized from 
\begin{eqnarray}
\rho_S^{\textrm{fact}}(0)&=&\rho_{S_1}\otimes\rho_{S_2}\otimes\cdots\otimes\rho_{S_N}
\label{Eqn:FctId}
\end{eqnarray}
with identical single-atom states
\begin{eqnarray}
\rho_{S_1}=\rho_{S_2}=\cdots=\rho_{S_N}=\left(
               \begin{array}{cc}
               \rho_{ee} & \rho_{eg} \\
               \rho_{eg}^* & \rho_{gg} \\
               \end{array}
               \right).
\label{Eqn:SingleAtom}
\end{eqnarray}
Such states are fully characterized by $\rho_{ee}$ and $\rho_{eg}$ and may also be superradiant. 
For example, states of this form are experimentally studied in Ref. \cite{Kim} 
to realize the single-atom superradiance where a serious of atoms enter a cavity one by one.
In Ref. \cite{Cheng}, we show that such a model can be used to achieve the  Heisenberg Limit 
in a quantum metrology scheme for measuring the atom-cavity coupling. The superradiance 
of two artificial atoms with initial state $\frac{1}{2}(\ket{e}+\ket{g})\otimes(\ket{e}+\ket{g})$ 
are experimentally observed in Ref. \cite{Mlynek}. 
In the above works, the single-atom coherence ($|\rho_{eg}|$) are 
crucial for creating superradiance.  Thus it is desirable to know the
influence of such states on the memory effects in superradiance, especially the 
role of the single-atom coherence.      

\begin{figure}
\includegraphics*[width=9cm]{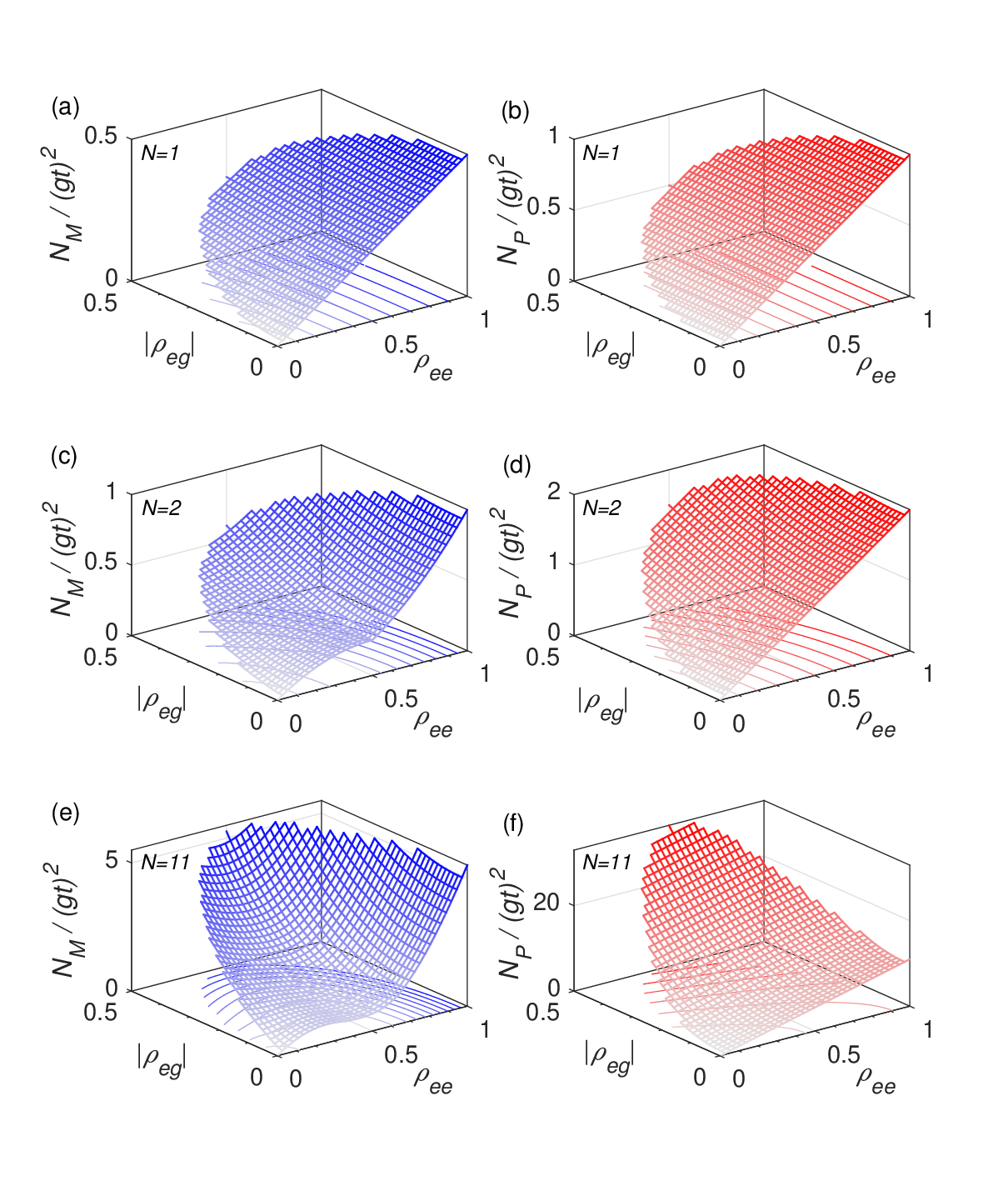}
\caption{Normalized memory effects $N_M$ (left column) and 
normalized cavity photon number $N_P$ (right column) for $\rho_S^{\textrm{fact}}(0)$ as functions 
of $\rho_{ee}$ and $|\rho_{eg}|$  in the early-time regime.  
The atom numbers are $N=1$, $2$, and  $11$ from top to bottom. 
 For any $\rho_{ee}$ and $N$ ($N>1$), $N_M$ ($N_P$) can always 
 be enhanced nonlinearly by the single-atom coherence $|\rho_{eg}|$.  }
\label{FIG:NmNpEeEg}
\end{figure}

In the following, the normalized value of memory effects and the normalized
cavity photon number for states described by Eq.\ (\ref{Eqn:FctId}) are calculated
using  Eqs.\ (\ref{Eqn:NMNorm}) and  (\ref{Eqn:NpNorm}).  When Eq.\ (\ref{Eqn:SingleAtom}) is
a pure state, i.e.,  $\rho_{Sn}=\ket{\phi}\bra{\phi}$, $\rho_S^{\textrm{fact}}(0)$ can be decomposed into the
superpositions of different Dicke states \cite{Kim}. Using the analytical result of 
$\bra{\phi} \sigma^+\sigma^-\ket{\phi}$ in Ref. \cite{Kim},  we obtain the normalized 
cavity photon number for $\rho_S^{\textrm{fact}}(0)$: 
\begin{eqnarray}
\frac{N_P[\rho_S^{\textrm{fact}}(0)]}{(gt)^2}=N(N-1)|\rho_{eg}|^2+N\rho_{ee}.
\label{Eqn:NpFct}
\end{eqnarray}

In contrast with Eq.\ (\ref{Eqn:NMJM}), the analytical solution of Eq.\ (\ref{Eqn:NMNorm}) 
for the factorized initial states is more complicated in general. Therefore, Eq.\ (\ref{Eqn:NMNorm}) 
is solved numerically here. The results of the normalized $N_M[\rho_S^{\textrm{fact}}(0)]$ and 
$N_P[\rho_S^{\textrm{fact}}(0)]$ are depicted in Fig.\ \ref{FIG:NmNpEeEg} as functions of $\rho_{ee}$ and 
$|\rho_{eg}|$ with $N=1$, 2 and $11$. The condition  $|\rho_{eg}|^2\leqslant(1-\rho_{ee})\rho_{ee}$  
are satisfied in Fig.\ \ref{FIG:NmNpEeEg}  to guarantee the semipositivity of  $\rho_{S_n}$. 
The single-atom coherence $\rho_{eg}$ are chosen to be real in the simulations for
simplicity. Numerical results  shows that the phase of $\rho_{eg}$ does not affect the 
value of memory effects.  Unlike the case for the Dicke states,  $N_M[\rho_S^{\textrm{fact}}(0)]/(gt)^2$ is 
not proportional  to $N_P[\rho_S^{\textrm{fact}}(0)]/(gt)^2$ in general. It is observed that 
the normalized value of memory effects can always be enhanced by $|\rho_{eg}|$ for any $N$. 
Similar situations happen for $N_P[\rho_S^{\textrm{fact}}(0)]/(gt)^2$ when $N>1$ as seen in
 Eq.\ (\ref{Eqn:NpFct}) and Fig.\ \ref{FIG:NmNpEeEg} (right column). 
Besides, as $N$ increases, $N_M[\rho_S^{\textrm{fact}}(0)]/(gt)^2$ grows more and more fast 
with $|\rho_{eg}|$, so does $N_P[\rho_S^{\textrm{fact}}(0)]/(gt)^2$. 
Thus we infer that for states with large $N$, the cavity photon number $N_P$ are dominated 
by $|\rho_{eg}|$, and the memory effects are dominated by $N_P$.
In Fig.\ \ref{FIG:NmNpEeEg}(c) and (e), it is found that the states around    
$\rho_{eg}=0$ and $\rho_{ee}=0.5$ tend to take lower values of memory effects compared 
with their neighboring states, especially when $N$ is large.  This might be intuitively 
interpreted by the fact that $\rho_{eg}=0$ and $\rho_{ee}=0.5$ leads to a maximally
 mixed state of the $N$-atom system.  Besides, Fig.\ \ref{FIG:NmNpEeEg} shows that 
the (change of) the cavity photon number is not the only factor to determine the
value of the memory effects. Other factors might include the (change of) 
coherence of the cavity state between the basis $\ket{n}$.

\begin{figure}
\includegraphics*[width=9cm]{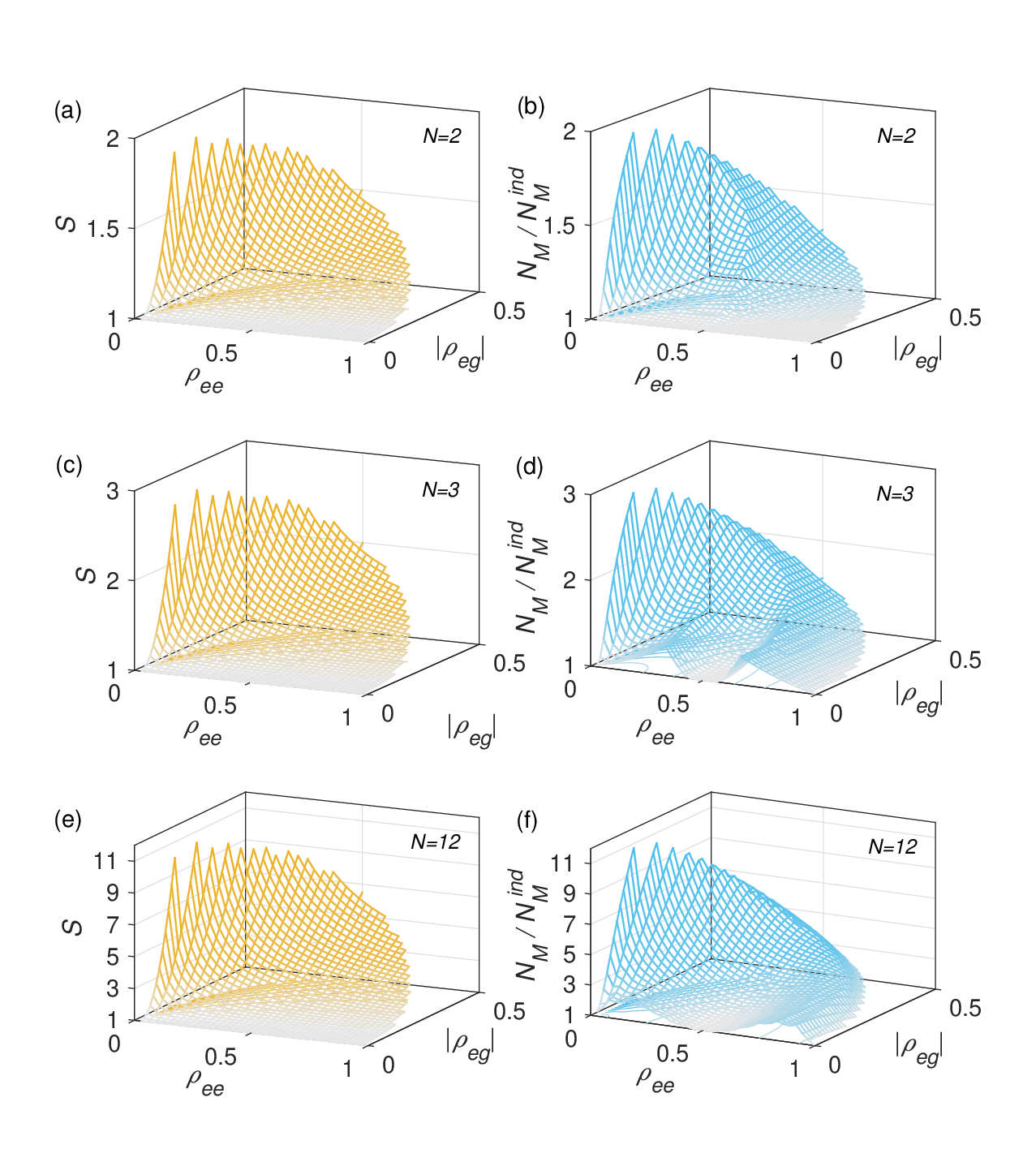}
\caption{
The degree of superradiance $S$ (left column) and the degree of memory-effect-enhancement  
$N_M/N_M^{\textrm{ind}}$ (from independent radiation to collective radiation, right column)
 for $\rho_S^{\textrm{fact}}(0)$ as functions of $\rho_{ee}$ and $|\rho_{eg}|$  in the early-time regime.  
 The values of $S$ and $N_M/N_M^{\textrm{ind}}$  for $\rho_{ee}=0$ are not defined and not shown. 
 The atom numbers are $N=2$, $3$, and  $12$ from top to bottom. It is observed that 
 large values of $S$ and $N_M/N_M^{\textrm{ind}}$ happen for low-excitation 
 states with high coherence.} 
\label{FIG:S_RNM_fact}
\end{figure}

 According to Eqs.\ (\ref{Eqn:DgrSR}) and  (\ref{Eqn:NpFct}), the degree of superradiance for 
$\rho_S^{\textrm{fact}}(0)$ is given by 
\begin{eqnarray}
S[\rho_S^{\textrm{fact}}(0)]&=&\frac{N_P[\rho_S^{\textrm{fact}}(0)]}{\sum_{n=1}^{N}N_P^{(n)}(\rho_{S_n})} \nonumber\\
              &=&\frac{N(N-1)|\rho_{eg}|^2+N\rho_{ee}}{N\rho_{ee}}\nonumber\\
              &=&1+(N-1)\frac{|\rho_{eg}|^2}{\rho_{ee}}
\label{Eqn:DgrSRFact}
\end{eqnarray}
for state with $\rho_{ee}>0$.  The state $\rho_S^{\textrm{fact}}(0)$ is superradiant when $|\rho_{eg}|$ is 
nonzero and $N>1$. Besides, the degree of superradiance grow quadratically with the single-atom 
coherence $|\rho_{eg}|$ for $N>1$. 
The maximum of $S[\rho_S^{\textrm{fact}}(0)]$ is $N$ when $\rho_{ee}\rightarrow0$ and  
$|\rho_{eg}|=\sqrt{(1-\rho_{ee})\rho_{ee}}$ (its maximum under the semi-positivity constraint). 
We illustrate the degrees of superradiance for different atom numbers in Fig.\ \ref{FIG:S_RNM_fact} (left column).
As done in last section, we also calculate the degree of memory-effect-enhancement  $N_M[\rho_S^{\textrm{fact}}(0)]/N_M^{\textrm{ind}}[\rho_S^{\textrm{fact}}(0)]$ in the early-time regime
with Eqs.\ (\ref{Eqn:NMNorm}) and (\ref{Eqn:NMNormInd}) for different atom numbers.
These results are shown in Fig.\ \ref{FIG:S_RNM_fact} (right column). Unlike the case for the Dicke states, 
the dependence of $N_M/N_M^{\textrm{ind}}$ on $\rho_{ee}$ and $\rho_{eg}$ is not the same 
as that of $S$. But there are many similarities between them. When there are only 
one atom ($N=1$), $N_M[\rho_S^{\textrm{fact}}(0)]/N_M^{\textrm{ind}}[\rho_S^{\textrm{fact}}(0)]=S[\rho_S^{\textrm{fact}}(0)]=1$ 
according to our definitions. For $N=2$, $S$ and $N_M[\rho_S^{\textrm{fact}}(0)]/N_M^{\textrm{ind}}[\rho_S^{\textrm{fact}}(0)]$ 
are slightly different, implying that the memory effects are not merely determined by the 
cavity photon numbers for the factorized identical states.  
As the atom number increases (for $N\geqslant3$), the surface shapes 
in the right-side figures become smoother and more similar to those in
the left column, especially when $S$ is large. As shown in Fig.\ \ref{FIG:S_RNM_fact}, 
strong memory-effect-enhancement happens for low-excitation states with high coherence (as $S$ does). 
Similar to Eq.\ (\ref{Eqn:DgrSRFact}), numerical simulations demonstrate 
that the maximum of $N_M[\rho_S^{\textrm{fact}}(0)]/N_M^{\textrm{ind}}[\rho_S^{\textrm{fact}}(0)]$ is also $N$ 
when $\rho_{ee}\rightarrow 0$ and $|\rho_{eg}|=\sqrt{(1-\rho_{ee})\rho_{ee}}$. 
These results imply that for states with high degree of superradiance, 
the (enhancement of) memory effects are dominated by the cavity photon number. 
Besides, numerical simulations show that for initial states whose 
spontaneous radiation can be enhanced by a common environment ($S>1$), 
their memory effects can always be enhanced simultaneously ($N_M[\rho_S^{\textrm{fact}}(0)]/N_M^{\textrm{ind}}[\rho_S^{\textrm{fact}}(0)]>1$).
Therefore, we infer that the memory effects also contribute to the establishment 
of the superradiance for the factorized identical initial states.

\subsection{Other initial states}
Entanglement and coherence of the initial states may play important roles in 
superradiance \cite{Agarwal,Tan,Wolfe,Tasgin,Lohof}, especially for the Dicke states.
Next, we examine the influence of the coherence (or the entanglement among the atoms) 
in the dephased Dicke states $\rho^{\textrm{deph}}_{S}(0)$ on the memory effects 
as well as the superradiance. The initial state we consider is given by 
\begin{eqnarray}
\rho^{\textrm{deph}}_{S}(0)=\lambda\ket{JM}\bra{JM} +(1-\lambda)\mathcal{D}(\ket{JM}\bra{JM}) 
\label{Eqn:DephDicke}
\end{eqnarray}
where $0\leqslant\lambda\leqslant1$. Here $\mathcal{D}$ turns $\ket{JM}\bra{JM}$ into
 a fully dephased state with vanishing off-diagonal elements in the basis 
$\ket{e(g),e(g),\cdots,e(g)}$.  Meanwhile, $\mathcal{D}(\ket{JM}\bra{JM})$ is a mixed 
state of separable states whose entanglement is zero. 
Thus $\lambda$ reflects the strength of entanglement or the value of
 coherence in $\rho^{\textrm{deph}}_{S}(0)$. With the help of numerical solutions,
 we find by mathematical induction that $\frac{N_M[\mathcal{D}(\ket{JM}\bra{JM})]}{(gt)^2}=\frac{1}{2}(J+M)$
and 
\begin{eqnarray}
\frac{N_M[\rho^{\textrm{deph}}_S(0)]}{(gt)^2}&=&\lambda\frac{N_M(\ket{JM}\bra{JM})}{(gt)^2} \nonumber\\ 
& &+(1-\lambda)\frac{N_M[\mathcal{D}(\ket{JM}\bra{JM})]}{(gt)^2}\nonumber \\
                                 &=& \frac{1}{2}(J+M)[(J-M)\lambda+1].
\label{Eqn:NMDephJM}
\end{eqnarray} 
It is seen that the normalized value of memory effects increases
linearly with $\lambda$ when $M\neq J,-J$. The normalized cavity photon number can be
 calculated by Eq.\ (\ref{Eqn:NpNorm}). We obtain 
 \begin{eqnarray}
\frac{N_P[\rho^{\textrm{deph}}_S(0)]}{(gt)^2}=&=&\lambda\frac{N_P(\ket{JM})}{(gt)^2} \nonumber\\ 
& &+(1-\lambda)\frac{N_P[\mathcal{D}(\ket{JM}\bra{JM})]}{(gt)^2}\nonumber \\
                                 &=&(J+M)[(J-M)\lambda+1].
\label{Eqn:NpDeph}
\end{eqnarray}
Similarly, $N_M[\rho^{\textrm{deph}}_S(0)]=\frac{1}{2}N_P[\rho^{\textrm{deph}}_S(0)]=\max_{\tau_{10},\tau_{21}}\Delta N_{ex}^{\textrm{atom}}[\rho^{\textrm{deph}}_S(0)]$ 
is satisfied for the dephased Dicke states in the early-time regime.

The degree of superradiance for the dephased Dicke states can be calculated from Eq.\ (\ref{Eqn:NpDeph}) by 
\begin{eqnarray}
S[\rho^{\textrm{deph}}_S(0)]&=&\frac{N_P[\rho^{\textrm{deph}}_S(0)]}{\sum_{n=1}^{J+M}N_P(\ket{e})} \nonumber\\
                   &=& \frac{(J+M)[(J-M)\lambda+1]}{J+M}\nonumber\\
                   &=& (J-M)\lambda+1    
\label{Eqn:DgrSRdeph}
\end{eqnarray}
for $J+M>0$. It is observed that the entanglement or coherence (represented by  $\lambda$) in
the depahsed Dicke state is necessary for superradiance. The degree of superradiance increases linearly with $\lambda$ for superradiant Dicke states.  Next we discuss the degree of memory-effect-enhancement caused by a common environment.  When the initial state is $\mathcal{D}(\ket{JM}\bra{JM})$ (fully dephased), we find that $N_M^{\textrm{ind}}[\mathcal{D}(\ket{JM}\bra{JM}]/(gt)^2=N_M^{\textrm{ind}}[\ket{JM}\bra{JM}]/(gt)^2=\frac{1}{2}(J+M)$. 
Moreover, for a partially dephased Dicke state described by Eq.\ (\ref{Eqn:DephDicke}), 
there is $N_M^{\textrm{ind}}[\rho^{\textrm{deph}}_{S}(0)]/(gt)^2=\frac{1}{2}(J+M)$ regardless of $\lambda$.
Therefore, the degree of memory-effect-enhancement for $\rho^{\textrm{deph}}_{S}(0)$ is given by 
\begin{eqnarray}
\frac{N_M[\rho^{\textrm{deph}}_{S}(0)]}{N_M^{\textrm{ind}}[\rho^{\textrm{deph}}_{S}(0)]}=(J-M)\lambda+1=S[\rho^{\textrm{deph}}_S(0)]
\label{Eqn:DmeeDeph}
\end{eqnarray}
as the case in Sec.\textrm{IV} A.

At the end of this section, we calculate the value of the normalized value of memory effects
of a probabilistic mixture of Dicke states $\rho_S^{\textrm{mix}}(0)=\sum_M p_M\ket{JM}\bra{JM}$ where 
$J=N/2$, $M=-J,-J+1,\cdots,J$, and  $\sum_M p_M=1$. Using similar treatments as 
done in Eq.\ (\ref{Eqn:NMJMprov}), we find that 
\begin{eqnarray}
\frac{N_M[\rho_S^{\textrm{mix}}(0)]}{(gt)^2}\!&=&\!\frac{1}{4}[\!\sum_{M=-J}^{J-1}\!|p_{M+1} f(M+1)-p_{M} f(M)|\nonumber\\
& &+p_{J}f(J)]
\label{Eqn:NMJMmix}
\end{eqnarray} 
where $f(X)=(J+X)(J-X+1)$. The result reduces to Eq.\ (\ref{Eqn:NMJM}) when $\rho_S^{\textrm{mix}}(0)$ is
a pure state $\ket{JM}\bra{JM}$.

\section{Results in other regimes}
\subsection{Near-Markovian regime}
  
In the previous sections, we investigate the superradiance and memory effect characteristics of  
our model in the early-time regime with  quadratic expressions. 
The early-time dynamics can be understood by setting $\gamma=0$ in Eq.\ (\ref{Eqn:MEI})
 and consider the dynamics in a short time interval ($gt\ll1$). The definition of memory effects 
 in Ref. \cite{Hou2015} reflects the quality of Markovian approximation in a quantum process. Correspondingly, 
the initial-state-dependent memory effect in this paper reflects the quality of 
Markovian approximation conditioned on a particular initial state, which could be understood by 
$1-N_M[\rho_S(0)]$. By our definition, $N_M[\rho_S(0)]$ is generally nonzero (although small) for nontrivial exact dynamics. Therefore, it could be used to investigate another interesting regime, i.e., the near-Markovian regime, by setting $\gamma/g\gg1$ and considering the exact dynamics in  $0<t<\infty$ or as least a longer time
interval far beyond the environment memory time $\tau_E=1/\gamma$.  In this section, 
for similar reasons, we consider the early-stage (far beyond $\tau_E$) dynamics in the 
near-Markovian regime where the system's state does not change so much. In this way, we  
stress the influence of the initial states on the superradiance and the memory effects,
and compare the results with those in the early-time regime. The specific scale of the
early-stage time interval will be discussed in the following.

\begin{figure}
\includegraphics*[width=9cm]{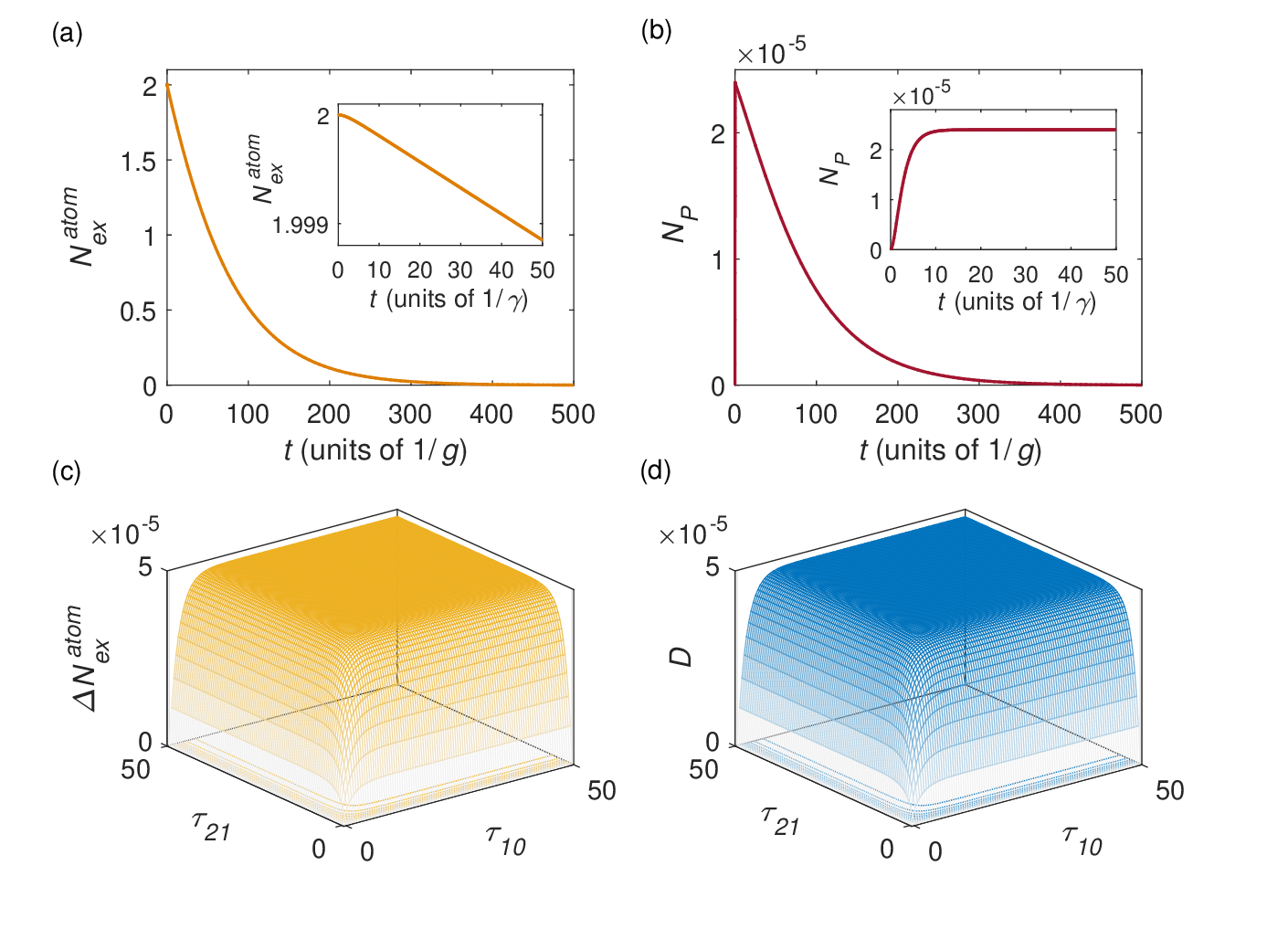}
\caption{
An example of the system and environment dynamics as well as the memory effect 
manifestations in a near-Markovian regime with $\gamma/g=10^3$ and a common cavity. The 
initial state is $\ket{JM}=\ket{2,0}$. (a) and (b) show the evolutions of 
the excitation number of the atoms $N_{ex}^{\textrm{atom}}$ and  the cavity photon number
$N_P$, respectively, where the insets display their early-stage dynamics.  
(c) and (d) show the difference of the excitation numbers
$\Delta N_{ex}^{\textrm{atom}}$ and the trace distance $D$, respectively, 
as functions of $\tau_{10}$ and $\tau_{21}$ (both in units of $1/\gamma$).} 
\label{FIG:NearMarkovian}
\end{figure}


The near-Markovian dynamics of the atoms can be calculated  by numerically
solving Eqs.\ (\ref{Eqn:MEI}) or (\ref{Eqn:MEIInd}) and tracing out the environment
under the condition  $\gamma/g\gg1$. We first present a near-Markovian  ($\gamma/g=10^3$)
example showing  typical system and environment dynamics and the memory effect manifestations 
in Fig.\ \ref{FIG:NearMarkovian}, where 4 atoms initially in the Dicke state $\ket{2,0}$
radiate in a common cavity. The system dynamics and environment dynamics are represented 
by the evolutions of the atom excitation number $N_{ex}^{\textrm{atom}}= \langle\sum_n\sigma_n^+\sigma_n^-\rangle$
and the cavity photon number $N_P=\langle b^\dag b\rangle$, respectively, as shown in 
Figs.\ \ref{FIG:NearMarkovian}(a) and \ref{FIG:NearMarkovian}((b). The memory effects are based on the trace 
distance of the final states in evolution A and C, i.e., $D=\frac{1}{2}\|\rho_S(t_2)-\rho'_S(t_2)\|$
which is shown in Fig.\ \ref{FIG:NearMarkovian}(d) as a function of $\tau_{10}=t_1-t_0$ and $\tau_{21}=t_2-t_1$.
Similarly, we also illustrate the difference of the atom excitation numbers denoted by 
$\Delta N_{ex}^{\textrm{atom}}=|\Tr[\sum_n\sigma_n^+\sigma_n^-\rho_S(t_2)]-\Tr[\sum_n\sigma_n^+\sigma_n^-\rho'_S(t_2)]|$
as a manifestation of memory effects in Fig.\ \ref{FIG:NearMarkovian}(c). 

 \begin{table*}[htbp]
 \caption{Characteristics of superradiance and memory effects for the dephased Dicke states
 ($\lambda=1$ for the Dicke states) in different regimes ($\tau_E=1/\gamma$).}
 \centering
 \begin{tabular}{c|c|c|c}
  \hline
  Characteristics  \quad &  \quad Early-time result \quad & \quad Characteristics  \quad &  \quad Early-stage Markovian result  \\
  \hline
    $N_P/(gt)^2$          &   $(J+M)[(J-M)\lambda+1]$             & $N_P^{\textrm{steady}}/(g\tau_E)^2$        & $4(J+M)[(J-M)\lambda+1]$ \\
   $N_M/(gt)^2$           & $\frac{1}{2}(J+M)[(J-M)\lambda+1]$    & $ N_M/(g\tau_E)^2$             & $8(J+M)[(J-M)\lambda+1]$ \\
   $N_P^{\textrm{ind}}/(gt)^2$     & $J+M$                                 & $N_P^{\textrm{steady},\textrm{ind}}/(g\tau_E)^2$    & $4(J+M)$ \\
   $N_M^{\textrm{ind}}/(gt)^2$     &   $\frac{1}{2}(J+M)$                  & $ N_M^{\textrm{ind}}/(g\tau_E)^2$       & $8(J+M)$ \\
       $S=N_P/N_P^{\textrm{ind}}$  & $(J-M)\lambda+1$                      &  $S=N_P^{\textrm{steady}}/N_P^{\textrm{stead},\textrm{ind}}$   & $(J-M)\lambda+1$ \\
    $N_M/N_M^{\textrm{ind}}$       &  $(J-M)\lambda+1$                     &  $N_M/N_M^{\textrm{ind}}$               & $(J-M)\lambda+1$ \\  
  \hline
 \end{tabular}
 \label{tab:SRNMEarlyMarDicke}
\end{table*}

\begin{table*}[htbp]
 \caption{Characteristics of superradiance and memory effects for the factorized identical states
 in different regimes ($\tau_E=1/\gamma$).}
 \centering
 \begin{tabular}{c|c|c|c}
  \hline
  Characteristics  \quad &  \quad Early-time result \quad & \quad Characteristics  \quad &  \quad  Early-stage Markovian result  \\
   \hline 
    $N_P/(gt)^2$        &   $N(N-1)|\rho_{eg}|^2+N\rho_{ee}$                                                            &  $N_P^{\textrm{steady}}/(g\tau_E)^2$        & $4[N(N-1)|\rho_{eg}|^2+N\rho_{ee}]$\\
   $N_M/(gt)^2$         & $\frac{1}{4}\|\mathcal{L}_{\sigma^-}[\rho_S(0)]\|$                                            &  $ N_M/(g\tau_E)^2$             & $4\|\mathcal{L}_{\sigma^-}[\rho_S(0)]\|$\\
   $N_P^{\textrm{ind}}/(gt)^2$   & $N\rho_{ee}$                                                                                  &  $N_P^{\textrm{steady},\textrm{ind}}/(g\tau_E)^2$    & $4N\rho_{ee}$\\
   $N_M^{\textrm{ind}}/(gt)^2$   &   $\frac{1}{4}\|\sum_{n=1}^N\mathcal{L}_{\sigma_n^-}[\rho_S(0)]\|$                            &   $ N_M^{\textrm{ind}}/(g\tau_E)^2$      & $4\|\sum_{n=1}^N\mathcal{L}_{\sigma_n^-}[\rho_S(0)]\|$\\
   $S=N_P/N_P^{\textrm{ind}}$    & $1+(N-1)|\rho_{eg}|^2/\rho_{ee}$                                                              &  $S=N_P^{\textrm{steady}}/N_P^{\textrm{stead},\textrm{ind}}$  & $1+(N-1)|\rho_{eg}|^2/\rho_{ee}$\\
    $N_M/N_M^{\textrm{ind}}$     &  $\|\mathcal{L}_{\sigma^-}[\rho_S(0)]\|/\|\sum_{n=1}^N\mathcal{L}_{\sigma_n^-}[\rho_S(0)]\|$  &   $N_M/N_M^{\textrm{ind}}$              & $\|\mathcal{L}_{\sigma^-}[\rho_S(0)]\|/   \|\sum_{n=1}^N\mathcal{L}_{\sigma_n^-}[\rho_S(0)]\|$ \\
    \hline
 \end{tabular}
 \label{tab:SRNMEarlyMarFact}
\end{table*}

As in Fig.\ \ref{FIG:NearMarkovian}, further numerical results shows that in a 
near-Markovian regime ($\gamma/g\gg1$) with initially excited Dicke states or factorized identical states,
the cavity photon number $N_P$ first experiences a rapid increase before $t\sim10/\gamma=10\tau_E$ 
and then reaches a plateau while the atom emission rate is almost constant as shown in the 
insets of Figs.\ \ref{FIG:NearMarkovian}(b) and \ref{FIG:NearMarkovian}(a).  After the plateau,
$N_P$ evolves on a larger timescale given by  $gt\propto\gamma/g$ where 
$N_P$ may monotonically decay for a highly superradiant initial state, or first increase and then decay,
for a less superradiant one. Thus the early stage of a near-Markovian dynamics in our work could
be understood as a time interval $[0,\tilde{t}]$ where $\tau_E \ll \tilde{t}\ll \tau_{L}$. 
Here $\tau_{L}$ is the lifetime of the radiation that is found to be proportional to $\gamma/g$
 and initial-state-dependent. In other words, in the early stage of a near-Markovian dynamics, 
 the environment photon number and the atom emission rate becomes almost steady 
 while the system does not change so much. Therefore, the 
almost-steady cavity photon number $N_P^{\textrm{steady}}$, as shown by the plateau in the inset of 
Fig.\ \ref{FIG:NearMarkovian}(b), represents the typical cavity photon number during the early stage 
of a near-Markovian process. In the near-Markovian regime, the memory effects $N_M[\rho_S(0)]$ can be calculated 
by Eq.\ (\ref{Eqn:NMI}) where $0\leqslant t_1\leqslant t_2\leqslant \tilde{t}$. Numerical 
evidences imply that there exists only one local maximum of $D=\frac{1}{2}\|\rho_S(t_2)-\rho'_S(t_2)\|$ 
when $\tau_{10}$ and $\tau_{21}$ are optimized (at least) in the early stage of a near-Markovian dynamics.
For example, in Fig.\ \ref{FIG:NearMarkovian}, the optimal $\tau_{10}$ and $\tau_{21}$ are
about $\tau_{10}=\tau_{21}\sim20/\gamma$ in $0\leqslant t_1\leqslant t_2\leqslant \infty$. 
In contrast, the optimal $\tau_{10}$ and $\tau_{21}$ that maximizing $\Delta N_{ex}^{\textrm{atom}}$
might be beyond the early-stage time interval for a less superadiant initial state. 
However, in practice, the values of $D$ as well as $\Delta N_{ex}^{\textrm{atom}}$ in the early stage of
a near-Markovian dynamics are nearly steady and robust against $\tau_{10}$ and $\tau_{21}$ 
after $\tau_{10}=\tau_{21}>10\tau_E$, especially when $\gamma/g$ is large. This makes it easy 
to define and numerically calculate the near-steady value of the excitation number difference 
$\Delta N_{ex,\textrm{steady}}^{\textrm{atom}}$.

We now discuss quantitatively the characteristics of the superadiance and the memory effects 
of the early-stage dynamics in near-Markovian regime. In Fig.\ \ref{FIG:NearMarkovian}, there 
are $N_M(\ket{2,0})=4.7954\times10^{-5}\approx4.8\times10^{-5}$ and
$N_P^{\textrm{steady}}(\ket{2,0})=R^{\textrm{steady}}(\ket{2,0})=2.3996\times10^{-5}\approx2.4\times10^{-5}$,
where $R=dN_{ex}^{\textrm{atom}}/dt$ is the photon emission rate of the atoms.
Meanwhile, the difference of cavity photon number $\Delta N_{ex}^{\textrm{atom}}(\tau_{10},\tau_{21})$ in Fig.\ \ref{FIG:NearMarkovian}(c)
 and the trace distance $D(\tau_{10},\tau_{21})$ in Fig.\ \ref{FIG:NearMarkovian}(d) are approximately the same. 
This implies that the difference of excitation numbers can be used to measure the memory effects for the Dicke 
states instead of the trace distance for simplicity. Further simulations imply that $N_M(\ket{2,0})\rightarrow8(g/\gamma)^2(J+M)[(J-M)+1]$
and $N_P^{\textrm{steady}}(\ket{2,0})=R^{\textrm{steady}}\rightarrow4(g/\gamma)^2(J+M)[(J-M)+1]$ as $\gamma/g$ goes larger 
and larger (the Markovian limit). For example, for $\gamma/g=10^4$, there are $N_M(\ket{2,0})=4.7999\times10^{-7}$
and $N_P^{\textrm{steady}}(\ket{2,0})=R^{\textrm{steady}}(\ket{2,0})=2.39999\times10^{-5}$ for the same time interval 
as in Fig.\ \ref{FIG:NearMarkovian}.

 Inspired by the above results, we conducted a number of simulations 
in the near-Markovian regime (with different $\gamma/g$) where the initial states 
included the (dephased) Dicke states with different $J$, $M$, and $\lambda$, 
and the factorized identical states with different $N$, $\rho_{ee}$, and $\rho_{eg}$.
As in former sections, we find that the characteristics of superradiance and memory effects in 
the early stage of a near-Markovian dynamics can be well approximated by analytic expressions. 
These characteristics include the steady cavity photon number of a common cavity ($N_P^{\textrm{steady}}$) 
and independent cavities ($N_P^{\textrm{steady},\textrm{ind}}=\langle\sum_{n}b_n^\dag b_n\rangle_{steady}$), 
the value of memory effects for a common cavity ($N_M$) and independent cavities ($N_M^{\textrm{ind}}$), 
the degree of superradiance $S=\langle \sigma^+\sigma^-\rangle/\sum_n\langle \sigma_n^+\sigma_n^-\rangle=N_P^{\textrm{steady}}/N_P^{\textrm{stead},\textrm{ind}}=R^{\textrm{steady}}/R^{\textrm{steady},\textrm{ind}}$,
and the degree of memory-effect-enhancement $N_M/N_M^{\textrm{ind}}$. Using $\tau_E$ instead of $1/\gamma$ 
for formal correspondence, these early-stage characteristics of the superradiance and the memory 
effects in the near-Markovian regime are listed in Table \ref{tab:SRNMEarlyMarDicke} 
(dephased Dicke states) and \ref{tab:SRNMEarlyMarFact} (factorized identical states) and compared with the early-time results. As shown in Table \ref{tab:SRNMEarlyMarDicke} and \ref{tab:SRNMEarlyMarFact}, the relations between the 
early-stage Markovian results and the early-time results can be summarized by 
\begin{eqnarray}
\frac{N_P^{\textrm{steady}}[\rho_S(0)]}{(g\tau_E)^2}&=&4\frac{N_P[\rho_S(0)]}{(gt)^2}, \\
\frac{N_M[\rho_S(0)]}{(g\tau_E)^2}&=&16\frac{N_M[\rho_S(0)]}{(gt)^2}
\label{Eqn:relations}
\end{eqnarray}
for both collective and independent radiations and the two types of initial states. 
Therefore, the degree of superradiance $S$ in the early stage of a Markovian regime 
is consistent with that in the early-time regime, so does the degree of memory-effect-enhancement $N_M/N_M^{\textrm{ind}}$. Besides, for the Dicke states, there are $\Delta N_{ex}^{\textrm{atom},\textrm{steady}}\approx2N_P^{\textrm{steady}}\approx N_M$ and 
$\Delta N_{ex}^{\textrm{atom}}(\tau_{10},\tau_{21})\approx D(\tau_{10},\tau_{21})$
in the early stage of the near-Markovian dynamics. In comparison, for the factorized identical states, 
the following relations holds approximately in the early stage of the near-Markovian dynamics:
 $\Delta N_{ex}^{\textrm{atom},\textrm{steady}}\approx2N_P^{\textrm{steady}}$ and $\Delta N_{ex}^{\textrm{atom}}(\tau_{10},\tau_{21})\propto D(\tau_{10},\tau_{21})$.  The results in this section demonstrate that the characteristics
of memory effects and superradiance in the early stage of a near-Markovian dynamics 
can be well reflected by its early-time dynamics.

\subsection{Strongly non-Markovian regime}

In this section, we continue to discuss the influence of the initial states on the superradiance
and memory effects in a strongly non-Markovian regime where $g$ and $\gamma$ is comparable and 
 $0\leqslant t <\infty$ are considered. In this regime, typically there are backflows of 
photons from the cavity to the atoms and the values of memory effects are much 
higher than those in the early-time regime or the near-Markovian regime. It is worth noting
that, in this regime, the superradiance characteristics, such as the
atom emission rate $R=dN_{ex}^{\textrm{atom}}/dt$ and the cavity photon number $N_P$, are neither quadratic nor
steady in time. Therefore, we may discuss their maxima (denoted by $R^{\textrm{max}}$ and $N_P^{\textrm{max}}$) and 
use $S=R^{\textrm{max}}/R^{\textrm{max},\textrm{ind}}$ as the degree of superradiance. Moreover (unlike what happens in other regimes), 
the initial states usually have evolved dramatically when the atom emission rate $R$ or other quantities
reach their maxima. 
 
\begin{figure}
\includegraphics*[width=9cm]{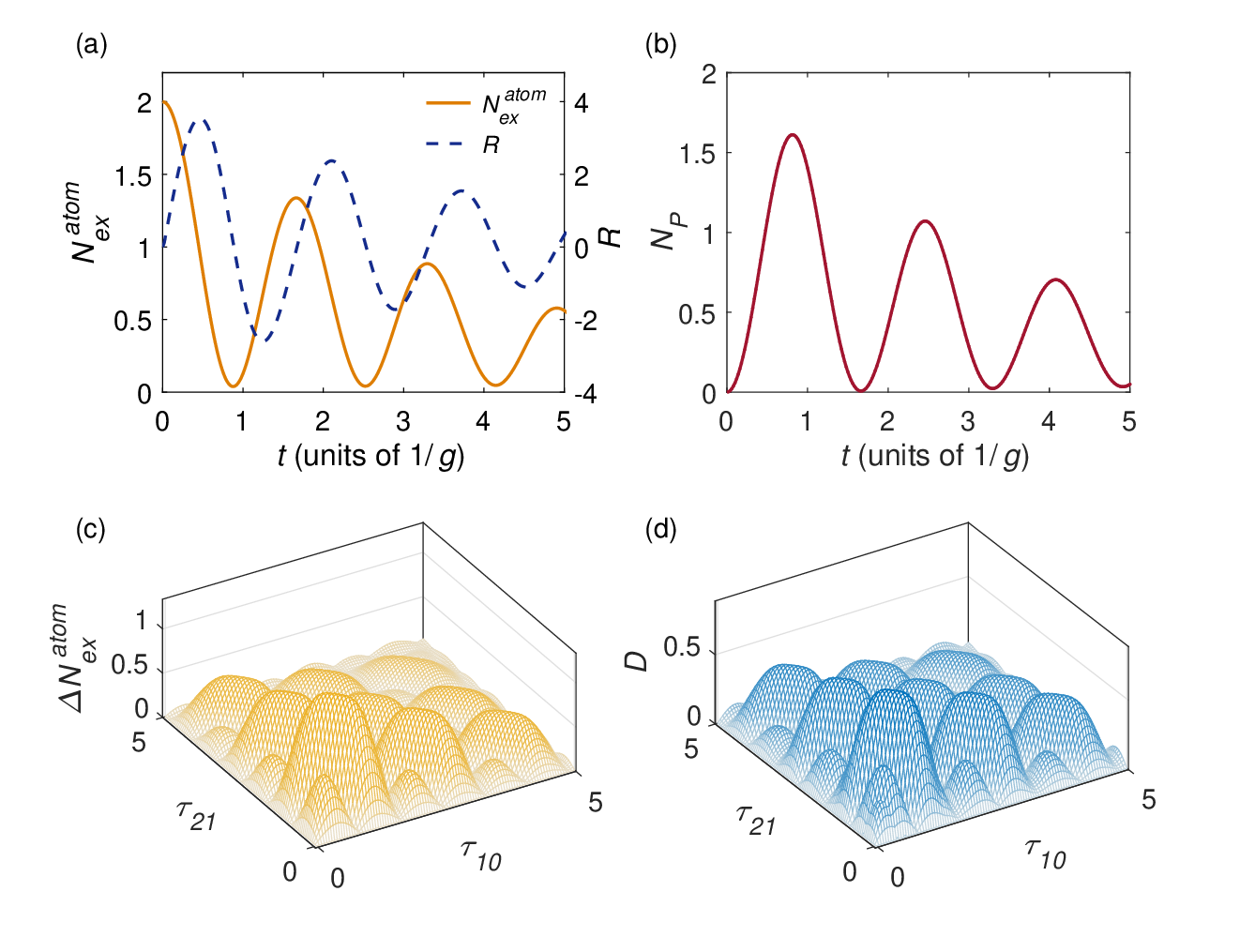}
\caption{
An example of the system and environment dynamics as well as the memory effect 
manifestations in a strongly non-Markovian regime with $\gamma/g=0.5$ and a common cavity. The 
initial state is $\ket{JM}=\ket{2,0}$. (a) shows the evolutions of the atom excitation number
 $N_{ex}^{\textrm{atom}}$ (solid line) and the emission rate of atoms $R=dN_{ex}^{\textrm{atom}}/dt$ (dashed line). 
 (b) shows the evolution of the cavity photon number $N_P$.  (c) and (d) show the
difference of the excitation numbers $\Delta N_{ex}^{\textrm{atom}}$ and the trace distance $D$, respectively, 
as functions of $\tau_{10}$ and $\tau_{21}$ (both in units of $1/g$).} 
\label{FIG:non-Markovianexample}
\end{figure}

As in last section, we provide a typical example in Fig.\ \ref{FIG:non-Markovianexample}
 showing the dynamics of the atoms and one common cavity as well as the memory effect manifestations 
 in a strongly non-Markovian regime with $\gamma/g=0.5$. The results are obtained by 
 numerically solving Eqs.\ (\ref{Eqn:MEI}) and (\ref{Eqn:MEIInd}). The initial state is the same as
 that in Fig.\ \ref{FIG:NearMarkovian}. As shown in Fig.\ \ref{FIG:non-Markovianexample}(a), there 
 are backflows of photons from the cavity to the atoms and the maximal atoms emission rate 
 happens during the first emission pulse. Besides, the maximal cavity photon number $N_P^{\textrm{max}}$ happens 
 at the end of the first emission pulse as shown in Fig.\ \ref{FIG:non-Markovianexample}(b).
 The trace distance $D=\frac{1}{2}\|\rho_S(t_2)-\rho'_S(t_2)\|$ as a function of 
 $\tau_{10}$ and $\tau_{21}$ are shown in Fig.\ \ref{FIG:non-Markovianexample}(d)
 where $N_M(\ket{2,0})=\max_{\tau_{10},\tau_{21}} D(\tau_{10},\tau_{21})\approx 0.89$, 
 implying that the dynamics is far from Markovian. Figure \ref{FIG:non-Markovianexample}(d)
 also demonstrate that the optimal $\tau_{10}$ and $\tau_{21}$ for the global maximum of 
 $D(\tau_{10},\tau_{21})$ can be found in a limited time interval. The manifestation of memory effects  
 by $\Delta N_{ex}^{\textrm{atom}}$ as a function of  $\tau_{10}$ and $\tau_{21}$ is shown 
 in Fig.\ \ref{FIG:non-Markovianexample}(c).  Although the surface shapes in 
 Figs.\ \ref{FIG:non-Markovianexample}(c) and  \ref{FIG:non-Markovianexample}(d) 
 are similar,  the value of $\Delta N_{ex}^{\textrm{atom}}$ 
 are not bounded as $D$ is. Further simulations with other initially excited Dicke states and
factorized identical states (with $\gamma/g=0.5$) show similar results as shown in 
Fig.\ \ref{FIG:non-Markovianexample}, e.g., there are backflows of photons and $R^{\textrm{max}}$ ($N_P^{\textrm{max}}$) 
happens during (after) the first emission pulse. Thus the degree of superradiance 
$S=R^{\textrm{max}}/R^{\textrm{max},\textrm{ind}}$ can be clearly defined in our simulations.

   \begin{figure}
\includegraphics*[width=9cm]{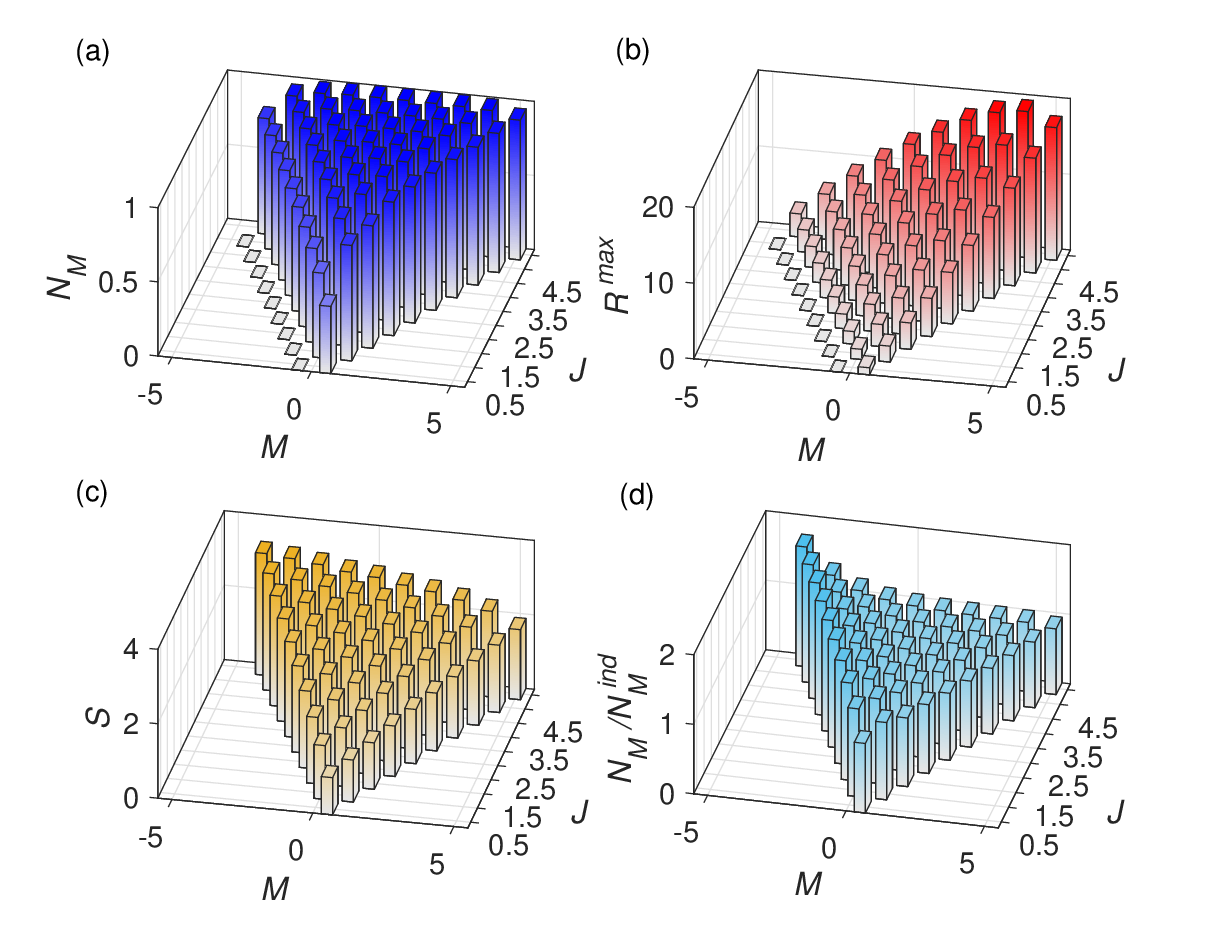}
\caption{
Some memory effect and superradiance characteristics in the strongly non-Markovian regime 
with $\gamma/g=0.5$ for the Dicke states. The number of atoms $N$ varies from 1 to 10, 
correspondingly, $J=1/2,3/2,\cdots,5$.  (a) and (b) show the value of memory effects $N_M$ 
and the maximal atoms emission rate $R^{\textrm{max}}$, respectively, as functions of $J$ and $M$.
Similarly, (c) and (d) show the degree of superradiance $S=R^{\textrm{max}}/R^{\textrm{max},\textrm{ind}}$ and the 
degree of the memory-effect-enhancement $N_M/N_M^{\textrm{ind}}$, respectively, where $M=-J+1,-J+2,\cdots,J$ } 
\label{FIG:JMnonMarkovian}
\end{figure}

We numerically calculated the value of memory effects $N_M$, 
the maximal atom emission rate $R^{\textrm{max}}$, the degree of superradiance $S$ and the degree of 
memory-effect-enhancement $N_M/N_M^{\textrm{ind}}$ for the Dicke states and the factorized identical states
with $\gamma/g=0.5$. The results for the Dicke states are shown in Fig.\ \ref{FIG:JMnonMarkovian} 
with no more than $10$ atoms. In the strongly non-Markovian regime, the values of memory effect
are close to 1 for most of the Dicke states as shown in Fig.\ \ref{FIG:JMnonMarkovian}(a), implying 
that the Markovian approximation are invalid for their dynamics. 
Unlike the results in the early-time regime or in the near-Markovian regime, $N_M(\ket{JM})$ is not 
proportional to $(J+M)(J-M+1)$ here. The first reason is that it tends to saturate 
($N_M[\rho_S(0)]\leqslant1$) for highly non-Markovian dynamics.  The second reason is 
as follows: on the timescale of the optimal $\tau_{10}$ ($\tau_{21}$) that maximizing 
$D(\tau_{10},\tau_{21})$, the initial states have evolved dramatically. Thus the 
connection between $N_M[\rho_S(0)]$ and $\rho_S(0)$ is weaker compared with those 
in the other two regimes we discussed.  
However, the dependence of $N_M(\ket{JM})$ on $J$ and $M$ is still roughly similar to that in
Fig.\ \ref{FIG:NmNpJM}. For example, for a fixed $J$, the strongest  memory effects happens for $M=0$ or
near 0. In the strongly non-Markovian regime, the radiation intensity is represented by the maximal 
emission rate $R^{\textrm{max}}$ as shown in Fig.\ \ref{FIG:non-Markovianexample}(a) (dashed line). 
For $J\leqslant5/2$, the state $\ket{J,J}$ have the strongest emission intensity with a given $J$. As $J$ increases, the states with the strongest emission intensity become $\ket{J,J-1}$ as shown in Fig.\ \ref{FIG:non-Markovianexample}(b).
We infer that a large $J$, the Dicke states with $M\sim0$ might be the most radiative ones.
Compared with the results in Fig.\ \ref{FIG:NmNpJM}, the Dicke states with small $J$ and large $M$ 
are more radiative than those with $M \sim 0$. The reason is as mentioned before, i.e., the 
states with large $M$ may have evolved to more superradiant ones when $R^{\textrm{max}}$ happens.
Figures \ref{FIG:JMnonMarkovian}(c) and \ref{FIG:JMnonMarkovian}(d) show the degree of superradiance and the degree of
memory-effect-enhancement. They are not proportional to $J-M+1$ as in Fig.\ \ref{FIG:DegreeNMComInd} 
in this regime. However, the highest degree of superradiance as well as memory-effect-enhancement 
in Figs.\ \ref{FIG:JMnonMarkovian}(c) and \ref{FIG:JMnonMarkovian}(d) still happens for the state with the largest $J$ and
$M=-J+1$. Other simulations show that, for the Dicke state $\ket{JM}=\ket{J,-J+1}$ (one excitation) in a common 
cavity, the memory effects for the dephased Dicke states satisfies $N_M[\rho^{\textrm{deph}}_S(0)]=\lambda N_M(\ket{JM}\bra{JM})+(1-\lambda)N_M[\mathcal{D}(\ket{J,M}\bra{JM})]$, i.e., $N_M[\rho^{\textrm{deph}}_S(0)]$ 
increases linearly with $\lambda$. For other Dicke states, this relation holds approximately at
 least for a few atoms.

\begin{figure}
\includegraphics*[width=9cm]{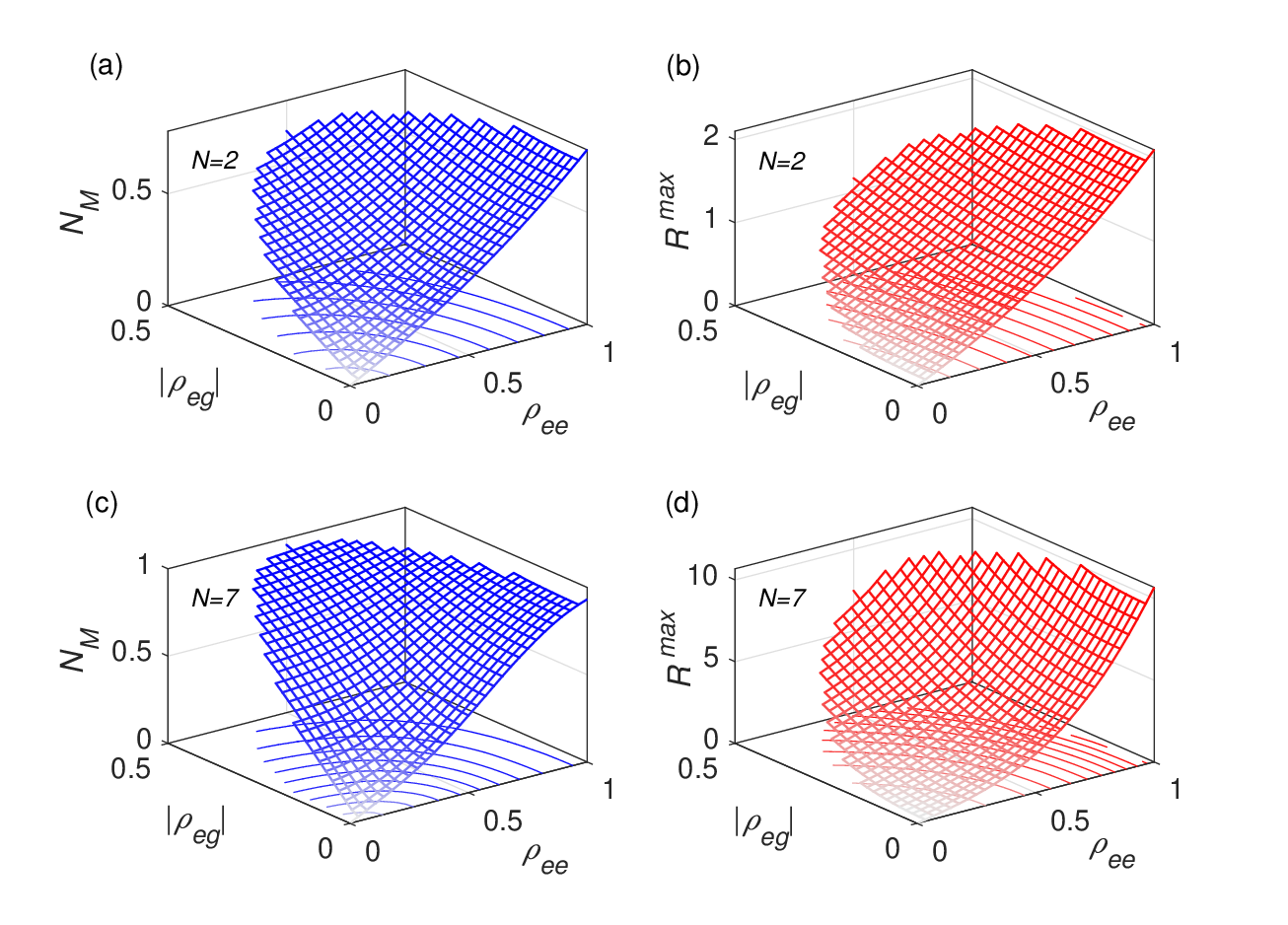}
\caption{
Memory effects $N_M$ (left column) and the maximal atom emission
 rate $R^{\textrm{max}}$ (right column) for  $\rho_S^{\textrm{fact}}(0)$ as 
 functions of $\rho_{ee}$ and $|\rho_{eg}|$ in the strongly non-Markovian 
 regime. The atom numbers are $N=2$ and $7$. } 
\label{FIG:NMRFactNonMarkovian}
\end{figure}

\begin{figure}
\includegraphics*[width=9cm]{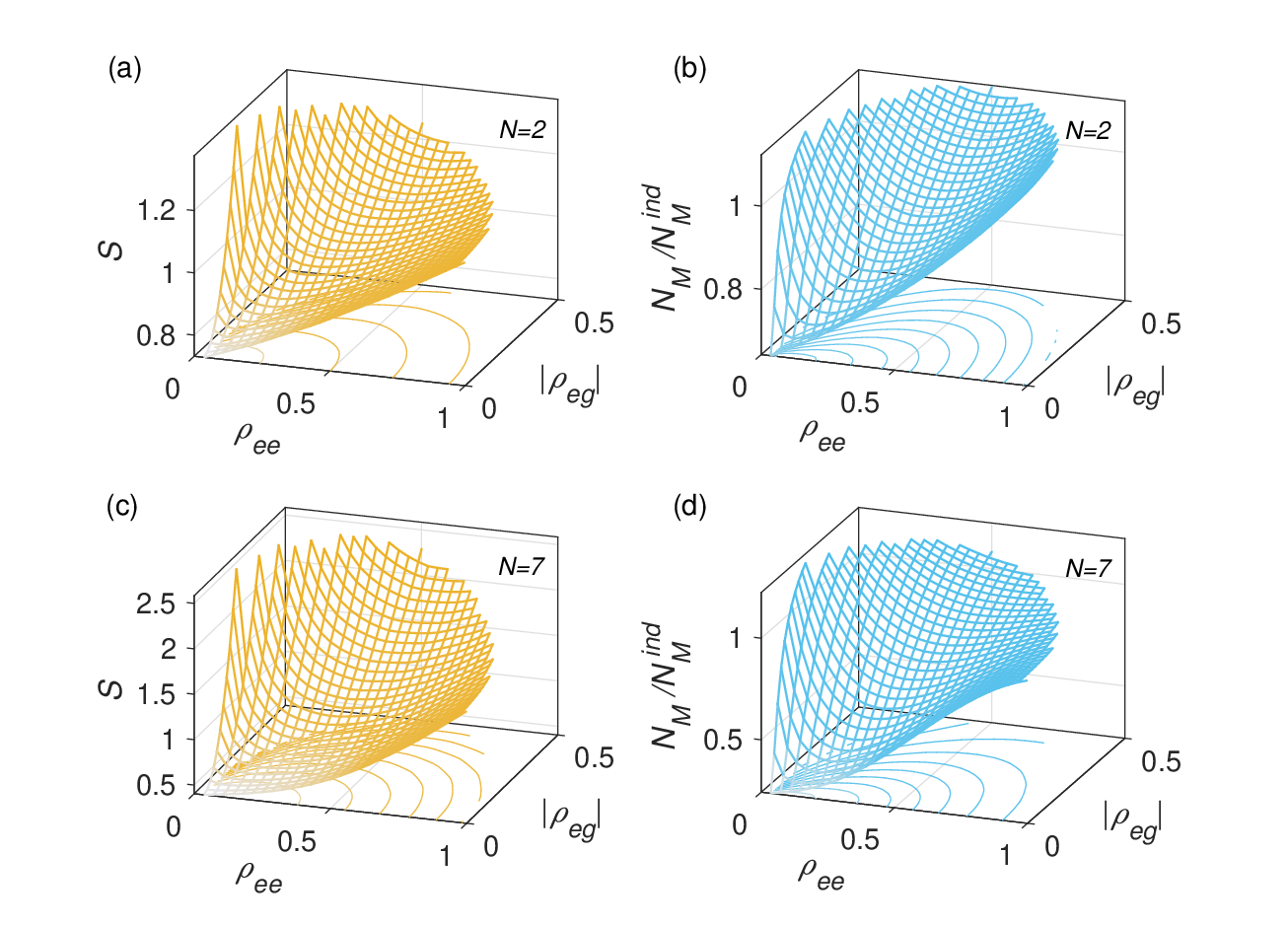}
\caption{
The degree of superradiance $S$ (left column) and the degree of memory-effect-enhancement  
$N_M/N_M^{\textrm{ind}}$ (right column) for  $\rho_S^{\textrm{fact}}(0)$ as functions of $\rho_{ee}$ and $|\rho_{eg}|$
in the strongly non-Markovian regime. The atom numbers are $N=2$ and $7$. The values of $S$ and 
$N_M/N_M^{\textrm{ind}}$  for $\rho_{ee}=0$ are not defined and not shown. } 
\label{FIG:SNMenhFactNonMarkovian}
\end{figure}

The value of memory effects  and the maximal atom emission rate for the factorized 
identical states are shown in Fig.\ \ref{FIG:NMRFactNonMarkovian} with two and seven atoms as 
functions of $\rho_{ee}$ and $|\rho_{eg}|$. In the strongly non-Markovian regime,
 the population $\rho_{ee}$, the coherence $|\rho_{eg}|$
and the atom number $N$ can increase the memory effects as well as the maximal atom emission rate. 
The results show some resemblance to those in Fig.\ \ref{FIG:NmNpEeEg} but with stronger memory 
effects and emission intensities in general. When $\rho_{ee}$ is small, the memory effects are 
still weak as the system almost stays in its ground state, which can be approximated by a Markovian  
dynamics in a trivial sense. The degree of superradiance $S=R^{\textrm{max}}/R^{\textrm{max},\textrm{ind}}$ and the degree of
memory-effect-enhancement $N_M/N_M^{\textrm{ind}}$ are shown in Fig.\ \ref{FIG:SNMenhFactNonMarkovian}. 
In contrast with the results in Fig.\ \ref{FIG:S_RNM_fact},
$S$ may be greater (less) than 1 for states with $\rho_{eg}=0$. Because when $R^{\textrm{max}}$ or $R^{\textrm{max},\textrm{ind}}$ 
happen, these initial states may evolve to more (less) superradiant ones. Similarly, $N_M/N_M^{\textrm{ind}}$ 
may also be less than 1 in the strongly non-Markovian regime. However, as in Fig.\ \ref{FIG:S_RNM_fact}, 
the coherence $|\rho_{eg}|$ can still increase the degree of superradiance, and initial states 
with low-excitations and strong coherence have large values of $S$ and $N_M/N_M^{\textrm{ind}}$. 
Besides, the surface shapes in Fig.\ \ref{FIG:SNMenhFactNonMarkovian}\ (right column) are roughly 
similar to those in Fig.\ \ref{FIG:SNMenhFactNonMarkovian}\ (left column).
Other calculations demonstrate that, for both $\ket{JM}$ and $\rho_S^{\textrm{fact}}(0)$ as 
initial states, the following relation holds roughly: $N_P^{\textrm{max}}\propto 
\Delta N_{ex}^{\textrm{atom},\textrm{max}} \propto R^{\textrm{max}}$ in our simulations.  Considering the 
weaker connection between the discussed characteristics and the initial states,
we may say that the results in the strongly non-Markovian regime are consistent
with those in the early-time regime.

\section{Discussions and Conclusions}
In this paper, we propose a method to evaluate the memory effects in a quantum process 
conditioned on a particular system initial state. The method is based on the physical 
interpretations of the inequality $T(t_2,t_0)\neq T(t_2,t_1)T(t_1,t_0)$ \cite{Hou2015}.
Some features of the non-Markovianity measure in Ref. \cite{Hou2015} are inherited. For example, 
nonzero memory effects can be characterized even in regimes where nonmonotonic behaviors 
do not occur, or have not occurred yet. Besides,  $0\leqslant N_M[\rho_S(t_0)] \leqslant 1$ 
is satisfied without additional normalizations. This allows us to compare the memory effects 
of quantum systems with different dimensions (e.g., different numbers of atoms in our model). 
Using our method, we calculate the influence of two types of initial states on the 
memory effects as well as the superradiance in different regimes. The characteristics of 
the memory effects and the superradiance are compared to reveal the role of memory 
effects in superradiance, and conversely, to understand the physical sources of the memory 
effects.

In the early-time (Zeno) regime and the (early-stage) near-Markovian regime where the 
initial states do not change so much, the main observations and conclusions are as follows. 
The memory effects for the (dephased) Dicke states are fully determined by the cavity photon number 
(connected to the atom radiation intensity) and they are proportional to $(gt)^2(J+M)[(J-M)\lambda+1]$ 
for quadratic dynamics, or $(g/\gamma)^2(J+M)[(J-M)\lambda+1]$ in the Markovian limit.
 Meanwhile, the degree of superradiance is equal to the degree memory-effect-enhancement 
 (from independent radiation to collective radiation) given by $(J-M)\lambda+1$. For the factorized identical states,
 the memory effects, the radiation intensity and the degree of superradiance can all be 
 enhanced by the single-atom coherence. The memory effects are closely related 
 to the cavity photon number, meanwhile, the superradiance ($S>1$) of an initial state 
is accompanied by the enhancement of memory effects (by a common cavity).  
These results demonstrate that the memory effects are closely related to 
the radiation intensity. Correspondingly, the enhancement of memory effects by 
a common environment plays an important role on the establishment of superradiance, 
especially for the Dicke states. On the other hand, these results imply that the (change of) cavity
photon number is one fundamental source of memory effects, especially for the Dicke states.
Besides, the entanglement in the Dicke states and the single-atom coherence in the 
factorized identical state are necessary for the existence of superradiance and 
important for the memory-effect-enhancement. 

For the whole dynamics in a strongly non-Markovian regime, an initial state 
may evolves dramatically during the radiation lifetime, making its influence weaker. 
However, simulations demonstrate that the relationships between the radiation intensity 
(degree of superradiance) and the memory effects (memory-effect-enhancement) are 
basically consistent with those in the other two regimes. The results in different 
regimes demonstrate that the characteristics of memory effects and superradiance 
in long-time dynamics can be reflected by its early-time behaviours.  In addition, 
our work also provides other ways to measure the memory effects through the atom 
excitation number or the cavity photon number.

\section{ACKNOWLEDGMENTS}
The authors thank X. L. Huang, X. L. Zhao, J. Cheng and  B. Cui  for 
helpful discussions. The work is supported by the National Natural Science Foundation
of China under Grant No. 11705026, the Fundamental Research Funds for the
 Central Universities in China under Grant No. 3132020178 and the 
 Natural Science Foundation of Liaoning Province under Grant No. 2023-MS-333.

\end{document}